\documentclass[twocolumn]{aastex631}

\let\splitbox\relax
\usepackage{comment}
\usepackage{enumitem}
\usepackage{verbatim}
\usepackage{amsmath}
\let\longtable*\relax
\usepackage{graphicx}
\usepackage{hyperref}
\usepackage{color}
\usepackage{xcolor}
\usepackage[export]{adjustbox}
\usepackage{overpic}
\usepackage[caption=false]{subfig}

\newcommand{\NH}{$N_\textrm{H}$\ }

\defcitealias{Ve22}{Ve22}
\defcitealias{Le22}{Le22}

\shorttitle{eROSITA bubbles are distant}
\shortauthors{Liu et al.}
\graphicspath{{./}{figures/}}

\turnoffedit
\turnoffeditone

\let\longtable*\relax
\begin{document}

\title{Morphological Evidence for the eROSITA Bubbles Being Giant and Distant Structures}

\author{Teng Liu}
%\email{liuteng@ustc.edu.cn}
\affiliation{Department of Astronomy, University of Science and Technology of China, Hefei 230026, China}
\affiliation{School of Astronomy and Space Science, University of Science and Technology of China, Hefei 230026, China}
\affiliation{Max-Planck-Institut für extraterrestrische Physik, Gießenbachstraße 1, 85748 Garching, Germany}
\author{Andrea Merloni}
\affiliation{Max-Planck-Institut für extraterrestrische Physik, Gießenbachstraße 1, 85748 Garching, Germany}
\author{Jeremy Sanders}
\affiliation{Max-Planck-Institut für extraterrestrische Physik, Gießenbachstraße 1, 85748 Garching, Germany}
\author{Gabriele Ponti}
\affiliation{INAF, Osservatorio Astronomico di Brera, Via E. Bianchi 46, 23807 Merate, (LC), Italy}
\affiliation{Max-Planck-Institut für extraterrestrische Physik, Gießenbachstraße 1, 85748 Garching, Germany}
\author{Andrew Strong}
\affiliation{Max-Planck-Institut für extraterrestrische Physik, Gießenbachstraße 1, 85748 Garching, Germany}
\author{Michael~C.~H. Yeung}
\affiliation{Max-Planck-Institut für extraterrestrische Physik, Gießenbachstraße 1, 85748 Garching, Germany}
\author{Nicola Locatelli}
\affiliation{INAF, Osservatorio Astronomico di Brera, Via E. Bianchi 46, 23807 Merate, (LC), Italy}
\author{Peter Predehl}
\affiliation{Max-Planck-Institut für extraterrestrische Physik, Gießenbachstraße 1, 85748 Garching, Germany}
\author{Xueying Zheng}
\affiliation{Max-Planck-Institut für extraterrestrische Physik, Gießenbachstraße 1, 85748 Garching, Germany}
\author{Manami Sasaki}
\affiliation{Dr. Karl Remeis-Sternwarte and Erlangen Centre for Astroparticle Physics, Friedrich-Alexander Universität Erlangen-Nürnberg, Sternwartstraße 7, 96049 Bamberg, Germany}
\author{Michael Freyberg}
\affiliation{Max-Planck-Institut für extraterrestrische Physik, Gießenbachstraße 1, 85748 Garching, Germany}
\author{Konrad Dennerl}
\affiliation{Max-Planck-Institut für extraterrestrische Physik, Gießenbachstraße 1, 85748 Garching, Germany}
\author{Werner Becker}
\affiliation{Max-Planck-Institut für extraterrestrische Physik, Gießenbachstraße 1, 85748 Garching, Germany}
\affiliation{Max-Planck-Institut für Radioastronomie Auf dem Hügel 69, 53121 Bonn, Germany}
\author{Kirpal Nandra}
\affiliation{Max-Planck-Institut für extraterrestrische Physik, Gießenbachstraße 1, 85748 Garching, Germany}
\author{Martin Mayer}
\affiliation{Dr. Karl Remeis-Sternwarte and Erlangen Centre for Astroparticle Physics, Friedrich-Alexander Universität Erlangen-Nürnberg, Sternwartstraße 7, 96049 Bamberg, Germany}
\author{Johannes Buchner}
\affiliation{Max-Planck-Institut für extraterrestrische Physik, Gießenbachstraße 1, 85748 Garching, Germany}

\begin{abstract}    
  There are two contradictory views of the eROSITA bubbles: either a $10^4$ pc-scale pair of giant bubbles blown by the Galactic center (GC), or a $10^2$ pc-scale local structure coincidentally located in the direction of GC.
  A key element of this controversy is the distance to the bubbles.
  Based on the 3D dust distribution in the Galactic plane, we found three isolated, distant (500--800 pc) clouds at intermediate Galactic latitudes. 
  Their projected morphologies perfectly match the X-ray shadows on the defining features of the north eROSITA bubble, i.e., the North Polar Spur (NPS) and the Lotus Petal Cloud (LPC), indicating that both the NPS and LPC are distant with a distance lower limit of nearly 1kpc.
  In the X-ray dark region between the NPS and LPC, we found a few polarized radio arcs and attributed them to the bubble's shock front. These arcs match up perfectly with the outer border of the NPS and LPC and provide a way to define the bubble's border.
The border defined in this way can be well described by the line-of-sight tangent of a 3D skewed cup model rooted in the GC. We conclude that, instead of being two independent, distant features, NPS and LPC compose a single, giant bubble, which, therefore, is most plausibly a 10-kpc scale bubble rooted at the GC.

\end{abstract}


\section{Introduction}
The eROSITA bubbles \citep{Predehl2020} are a huge 2D structure in the soft X-ray sky, spanning the Galactic longitude range of -60$\sim40\arcdeg$ and extending up to a latitude of about $80\arcdeg$.
Their most prominent part is the North Polar Spur (NPS), the arch-like feature located in the northeast ($l<180\arcdeg$) quadrant of the sky, which is loosely associated with Loop I \citep{Large1966}, the most prominent structure in the radio sky.
The 3D structure of the eROSITA bubbles is unknown.
Based on their four-fold symmetry around the Galactic center (GC) and the tantalizing association with the Fermi bubbles, they appear as 10$^4$ pc-scale  bubbles rooted in and blown by the GC \citep[e.g.,][]{Predehl2020,LaRocca2020}.
Alternatively, they could be a local structure at a hundred-pc scale \citep[e.g.][]{West2021,Das2020}, based on various pieces of evidence against the distant GC-blown model, including the alignment of polarization between NPS and nearby dust \citep[e.g.,][]{Panopoulou2021}, the association of NPS with nearby magnetic field \citep[e.g.,][]{Santos2021} and nearby neutral hydrogen \citep[e.g.,][]{Welsh2009}, and the nondetection of Faraday rotation associated with NPS \citep[e.g.,][]{Hutschenreuter2022}.
See \citet{Lallement22} and \citet{Sarkar2024} for extended reviews of the NPS and the eROSITA bubbles.

\begin{figure*}[t]
  \centering
  \parbox{0.45\textwidth}{
    \includegraphics[width=0.45\textwidth,right]{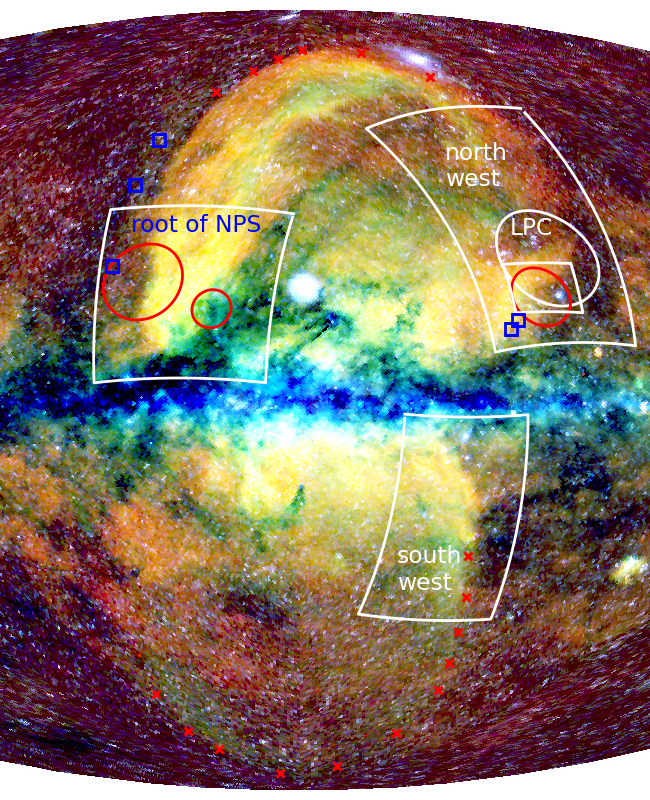}
    %\begin{overpic}[scale=0.49]{NEB_erass1_RGB.png}
    %\put(75,95){\color{red}\Large(a)}
    %\end{overpic}
    }
  \parbox{0.54\textwidth}{
    \includegraphics[width=0.51\textwidth,right]{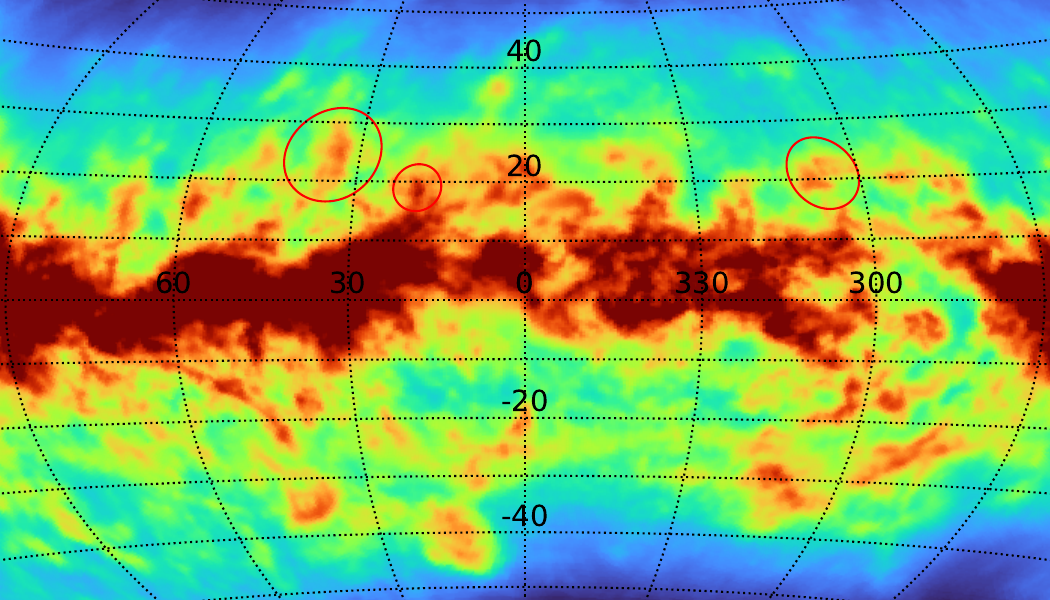}
    \begin{overpic}[scale=0.54]{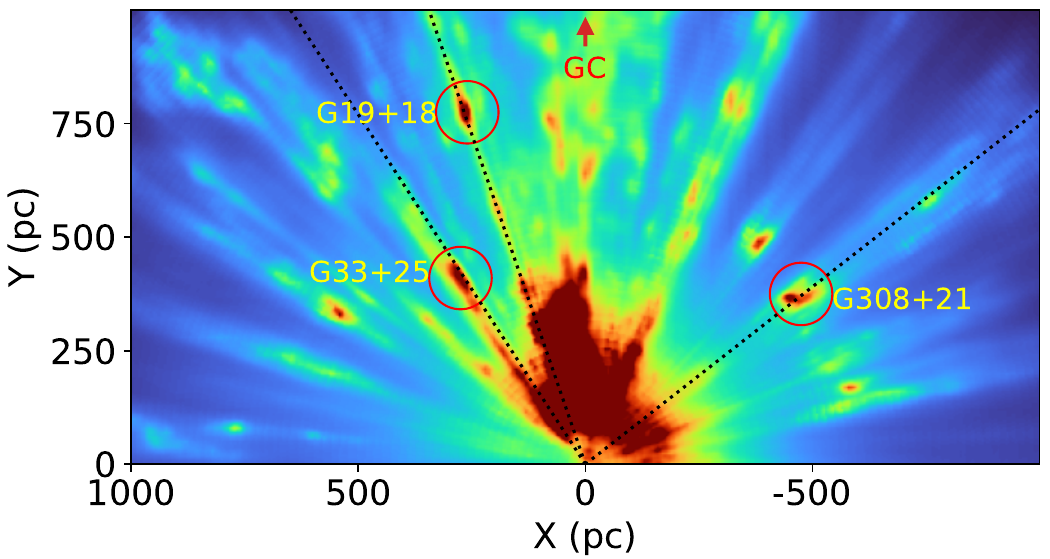}
    \put(-5,102){\color{red}\Large(a)}
    \put(5,48){\color{red}\Large(c)}
    \put(6,102){\color{red}\Large(b)}
    \end{overpic}
    }
  \caption{
    Panel (a) displays the eROSITA X-ray RGB image \citep{Predehl2020}, in which the white circle marks the LPC and the four white boxes are the image frames of Fig.~\ref{fig:LP_images} (G308+21), Fig.~\ref{fig:NPS_images} (NPS root), and Fig.~\ref{fig:X_I_U_Q} (northwestern border and southwestern bright rim).
    The red cross points and blue square points are manually selected marks of the X-ray bubble border (\S~\ref{sec:border}).
    Panel (b) display the dust extinction map in the 400--1000pc distance range of the \citetalias{Ve22} dust cube plotted in Hammer-aitoff projection.
    Panel (c) displays the extinction map of the pixels with Galactic latitude $>15\arcdeg$ in the \citetalias{Ve22} cube viewed from the north Galactic pole.
    In panel (c), the Sun is at (0,0), the direction of GC is to the top, and the dotted lines correspond to Galactic longitudes 33\arcdeg, 19\arcdeg, and 308\arcdeg.
    The three isolated clouds G19+18, G33+25, and G308+21 are marked on the three maps with red circles.
    \label{fig:Fig1}}
\end{figure*}

As displayed in Fig.~\ref{fig:Fig1} \citep{Predehl2020}, the North Polar Spur extends from the eastern ($l<180\arcdeg$) to the western Galactic hemisphere at high Galactic latitudes and fades in the latitude range 40$\sim 50\arcdeg$.
Below this X-ray dark region, there is an X-ray bright region ($295\arcdeg<l<310\arcdeg$, $23\arcdeg<b<33\arcdeg$), which has a regular shape that looks like a lotus petal (marked with a white circle in Fig.~\ref{fig:Fig1}). We set apart this structure for the first time, and name it the Lotus Petal Cloud (LPC).
We believe that the nature of LPC is a crucial piece of the puzzle in understanding the eROSITA bubbles.
If it shares the same origin as the NPS, it comprises the bubble's border in the western hemisphere and manifests a bubble shape that is highly asymmetric around $l=0\arcdeg$.
If, on the other hand, the LPC is a separate structure independent of the NPS, the morphological/symmetry argument of NPS composing a bubble would not hold, and the origin of LPC itself, as one of the brightest large structures in the sky, opens a new mystery that needs to be solved.
Thanks to the high-quality X-ray image obtained by eROSITA aboard the Spektrum Roentgen Gamma (SRG) satellite \citep{Predehl2021,Merloni2024}, we can see a sharp and round outer (western) border of the LPC.
The roundness of this border, which is a defining feature of LPC as an individual structure, provides a morphological argument that LPC is part of a bubble.

Over 70\% supernova remnants (SNRs) in the Milky Way have shell-type morphology characterized by a limb-brightened rim or shell at the location of the shock front, which usually coincides with the onset of X-ray emissions coming from the interior of the bubble. Radio rims represent polarized synchrotron emissions on the shock front, where particles are accelerated and the magnetic field is compressed with possibly other mechanisms in action to amplify it \citep[see reviews by][]{Reynolds2012,Dubner2017}. 
If a giant bubble is blown up by a powerful engine in the GC or the Galactic plane, the power could also create a shock front, similar to that commonly seen at the peripheries of SNRs.
The morphology of X-ray emission depends sensitively on the distribution of the gas that fills the bubble, which might be highly nonuniform. The shock front itself could also be affected by the ambient gas distribution, but with a thin rim pattern, its connection to the central engine is more straightforward.
In this work, we focus on the shape of the eROSITA bubbles' shock front rather than the X-ray brightness profile.

The distance to the NPS has been estimated using various methods, including the Faraday rotation of NPS \citep{Sun2015}, polarization alignment of NPS, starlight, and local dust \citep{Panopoulou2021}, and interstellar extinction \citep{Das2020,puspitarini2014}.
All these works above found that the high-latitude part of NPS being local ($\sim 100$pc or $200$pc).
Instead of the high-latitude part, some works studied the X-ray dark region below the root of NPS ($b<10\arcdeg$) \citep{Sofue2015,Lallement2016,Das2020}.
\citet{Lallement2016} confirmed that the southern terminus of NPS is absorption-bounded, and thus the NPS should extend down to a lower latitude, although it completely disappears at $b\sim 10\arcdeg$ in the eROSITA X-ray map. 
\citet{Das2020} analyzed the dust distribution in this $b<11\arcdeg$ region and concluded that the absorption towards the NPS is located within 700pc.
In this work, we provide a lower limit to the distance to the northern eROSITA bubble through isolated dusty clouds that obscure the bubble's X-ray emission. We focus on the most prominent X-ray features at intermediate-high latitudes (NPS and LPC), which are the defining features of the eROSITA bubble. The X-ray emission at low latitudes is less straightforward in inferring the distance to the eROSITA bubble, because there is so much material near the Galactic plane in the GC direction that the X-ray emission itself might be attributed to the background or back side of the bubbles.

In this paper, we argue that the eROSITA bubbles are indeed giant bubbles originating from the GC region, based on two arguments. In \S~\ref{sec:clouds}, we report the finding of a few distant, isolated, dusty clouds obscuring the NPS or the LPC and providing robust lower limits on their distances.
In \S~\ref{sec:shape}, we report the finding of a possible shock front traced by polarized radio arcs and use this to define the border of the eROSITA bubbles. We also present a 3D cup model to describe the border shape.
We discuss our results in \S~\ref{sec:discussion} and summarize the conclusion in \S~\ref{sec:conclusion}.

\section{Clouds Obscuring the northern eROSITA Bubble}
\subsection{Three Isolated Clouds}
\label{sec:clouds}
Based on a huge number of stars observed by Gaia and 2MASS, \citet[][]{La22} and \citet[][hereafter referred to as Ve22]{Ve22} presented the Galactic dust extinction density within a cube of the Galactic plane centered at the Sun.
Similarly, \citet{Leike2020} and \citet[][hereafter referred to as Le22]{Le22} also calculated such a dust cube using a different method.
From these 3D dust distributions, we \edit1{identified} a few isolated, distant clouds in front of the northern eROSITA bubble.
%From their high-resolution dust cube with a size of $3000\times 3000\times 800$pc \citepalias{Ve22},
Fig.~\ref{fig:Fig1} displays the dust distribution in two viewing aspects, the projected extinction map integrated in the distance range 400--1000pc and the 2D dust distribution parallel to the Galactic plane summing only the cube cells with $b>15\arcdeg$.
We selected three clouds, G19+18, G33+25, and G308+21, which are well isolated in the 3D space, and named them after their Galactic coordinates $l,b$ in degrees.
The first two are located around the root of the NPS and the third one is near the LPC.

Performing the Gaussian process regression in hundreds of $12.5\arcdeg\times 12.5\arcdeg$ patches across the sky, \citetalias{Le22} achieved relatively higher spatial resolutions than \citetalias{Ve22} in each patch but leaving discontinuities at the boundary of the patches.
In the \citetalias{Ve22} dust cube, the three clouds are also clearly shown and well isolated in the 3D space.
Compared to \citetalias{Ve22}, the distances of the clouds measured by \citetalias{Le22} are slightly lower and have smaller uncertainties.
More details of these clouds are discussed in the following two sections.

\citet{Willingale2013} presented an empirical relation to calculate the total hydrogen column density \NH by summing $N_\textrm{\footnotesize{H}\scriptsize{I}}$ and $N_\textrm{\footnotesize{H}\scriptsize{II}}$.
At the positions of the clouds, we used this relation to create the total \NH map, combining the HI4PI $N_\textrm{\footnotesize{H}\scriptsize{I}}$ map \citep{HI4PI2016} and the SFD extinction ($N_\textrm{\footnotesize{H}\scriptsize{II}}$) map \citep{SFD}.
\subsection{The distance to the LPC}
\begin{figure*}[bhpt]
%\centering
%  \begin{subfigure}{0.49\textwidth}
%\subfloat[plot]{}
\parbox{0.49\textwidth}{
  \begin{overpic}[scale=0.20]{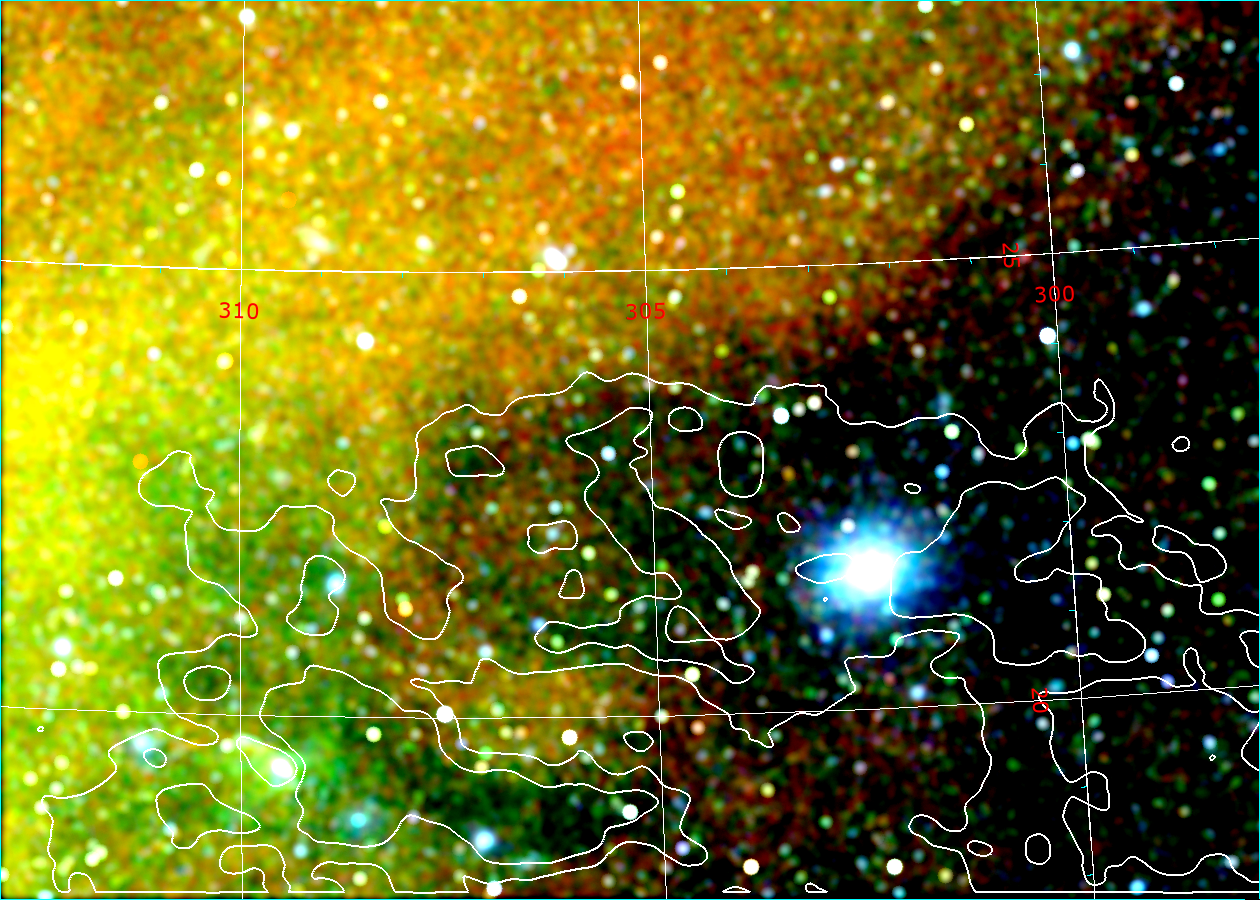}
    \put(90,5){\color{white}\large(a)}
  \end{overpic}
}
\vspace{-0.25cm}\\
\parbox{0.49\textwidth}{
  \begin{overpic}[scale=0.20]{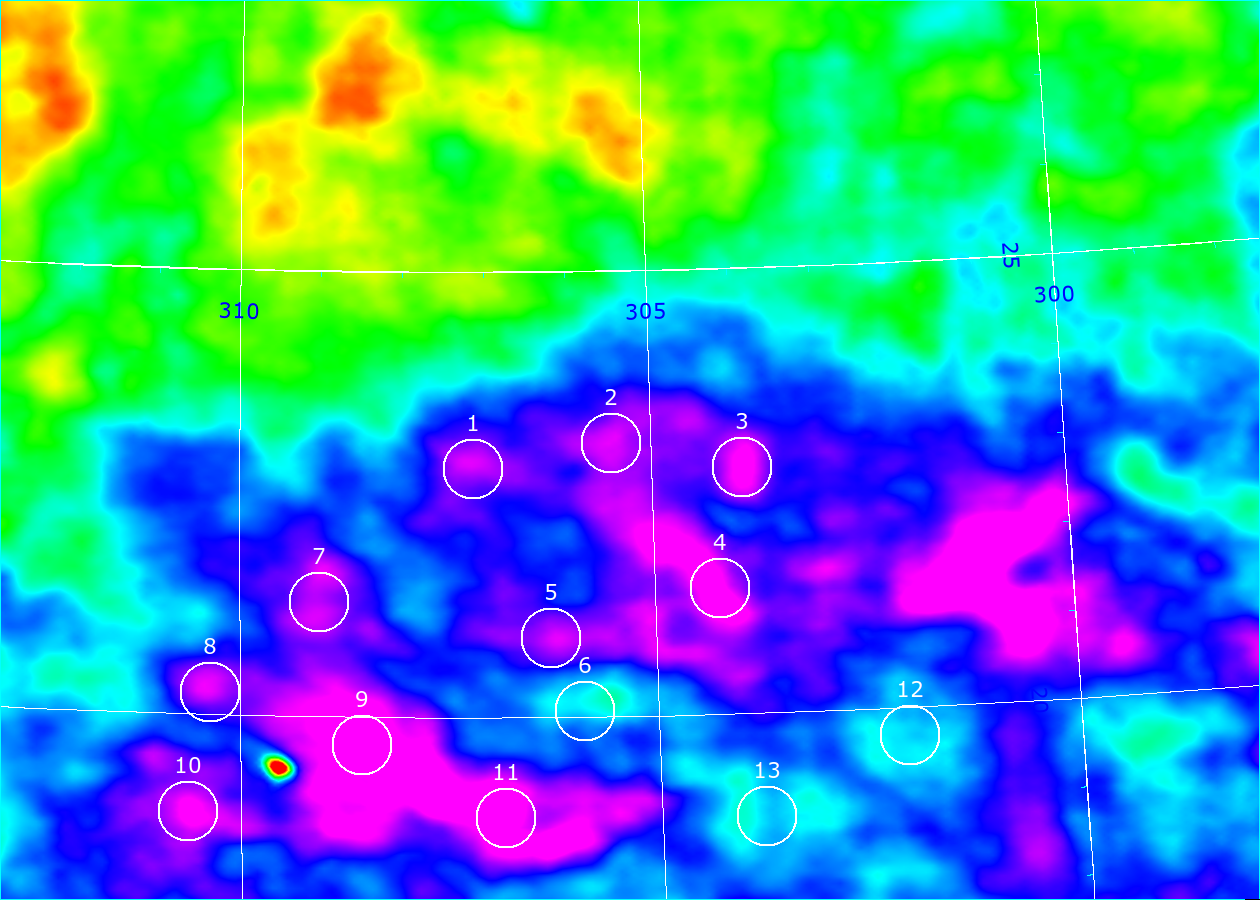}
    \put(90,65){\color{red}\large(b)}
  \end{overpic}
}
\parbox{0.49\textwidth}{
  \begin{overpic}[scale=0.20]{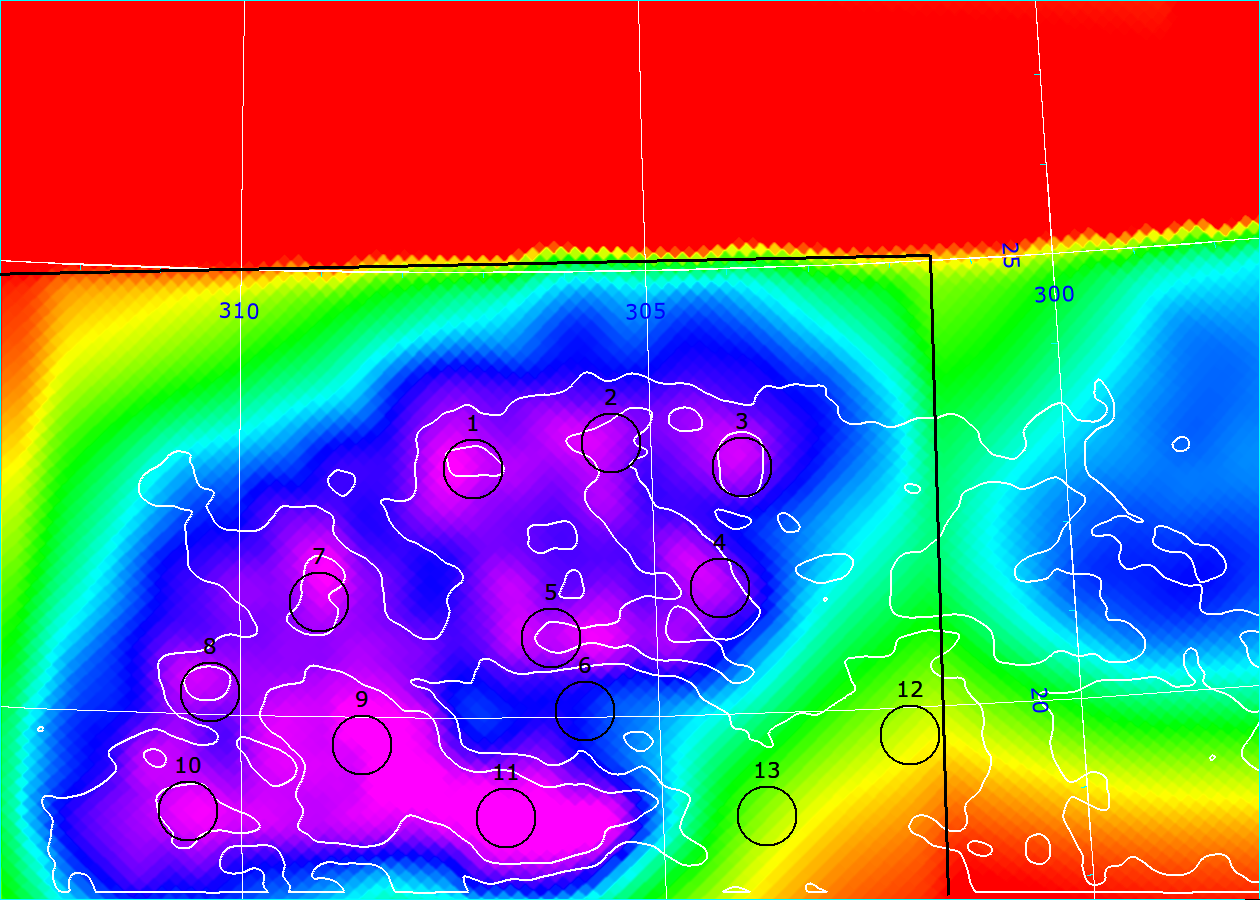}
    \put(5,40){\color{red}\large(c)}
    \put(4,100){\includegraphics[width=0.46\textwidth]{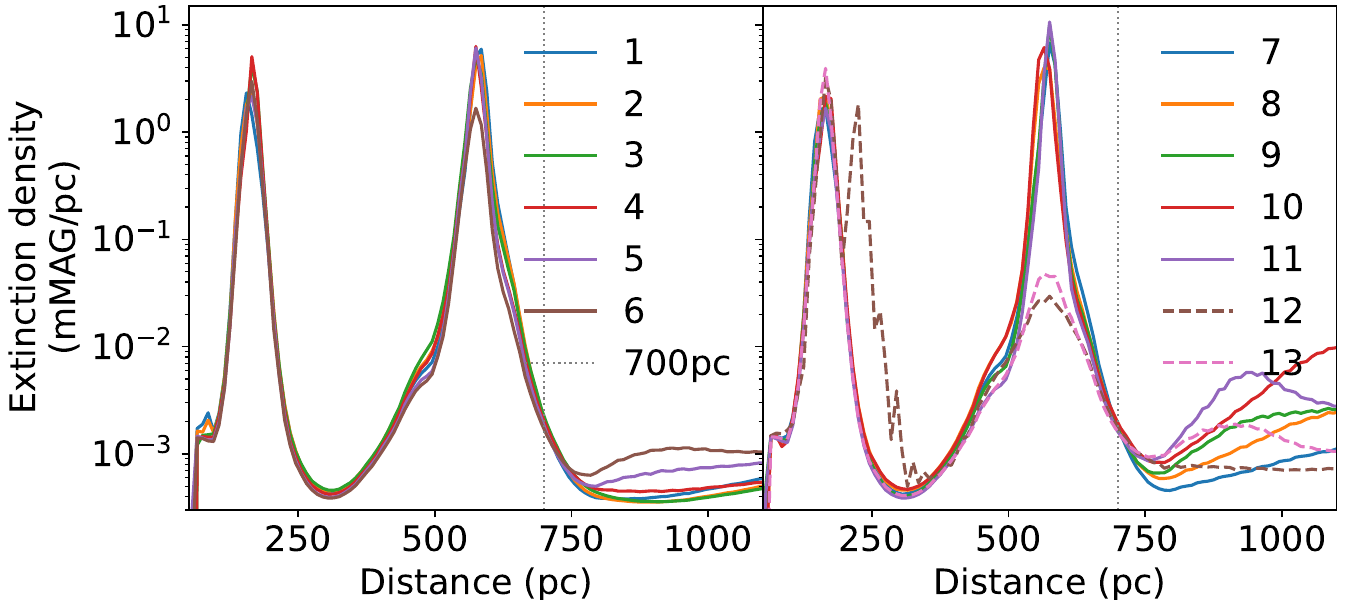}}
    \put(45,142){Leike+2022}
    \put(4,55){\includegraphics[width=0.46\textwidth]{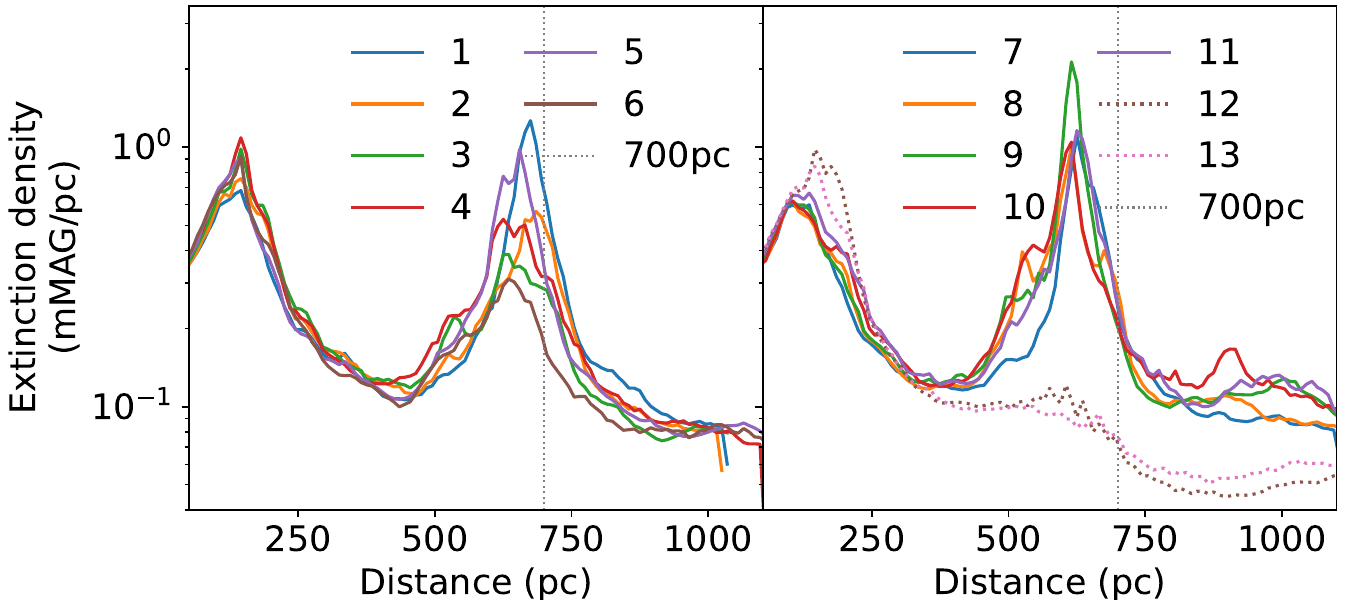}}
    \put(44,97){Vergely+2022}
    \put(6,98){\large (d)}
  \end{overpic}
  }
\caption{
  Panels (a), (b) and (c) display the eROSITA X-ray RGB image (R: 0.2--0.5~keV, G: 0.5--1~keV, B: 1--2~keV), the total \NH map, and the \citetalias{Le22} dust extinction map in the distance range 520--650 pc (with the patch border marked with black lines), respectively.
  %All images are plotted in the Equatorial system with a center of RA=195.633\arcdeg, DEC=-39.827\arcdeg, and a size of 14$\times$10\arcdeg.
  Both the extinction and \NH maps use a rainbow color map, where purple indicates large values.
  The contour of the total \NH map is overplotted on the X-ray image and the extinction map.
  $13$ positions are marked in the lower panels with 40\arcmin-diameter circles.
  Panel (d) displays the \citetalias{Le22} and \citetalias{Ve22} distance profiles of extinction at these positions, separating the positions into two groups for the sake of clarity.
 \label{fig:LP_images}}
\end{figure*}
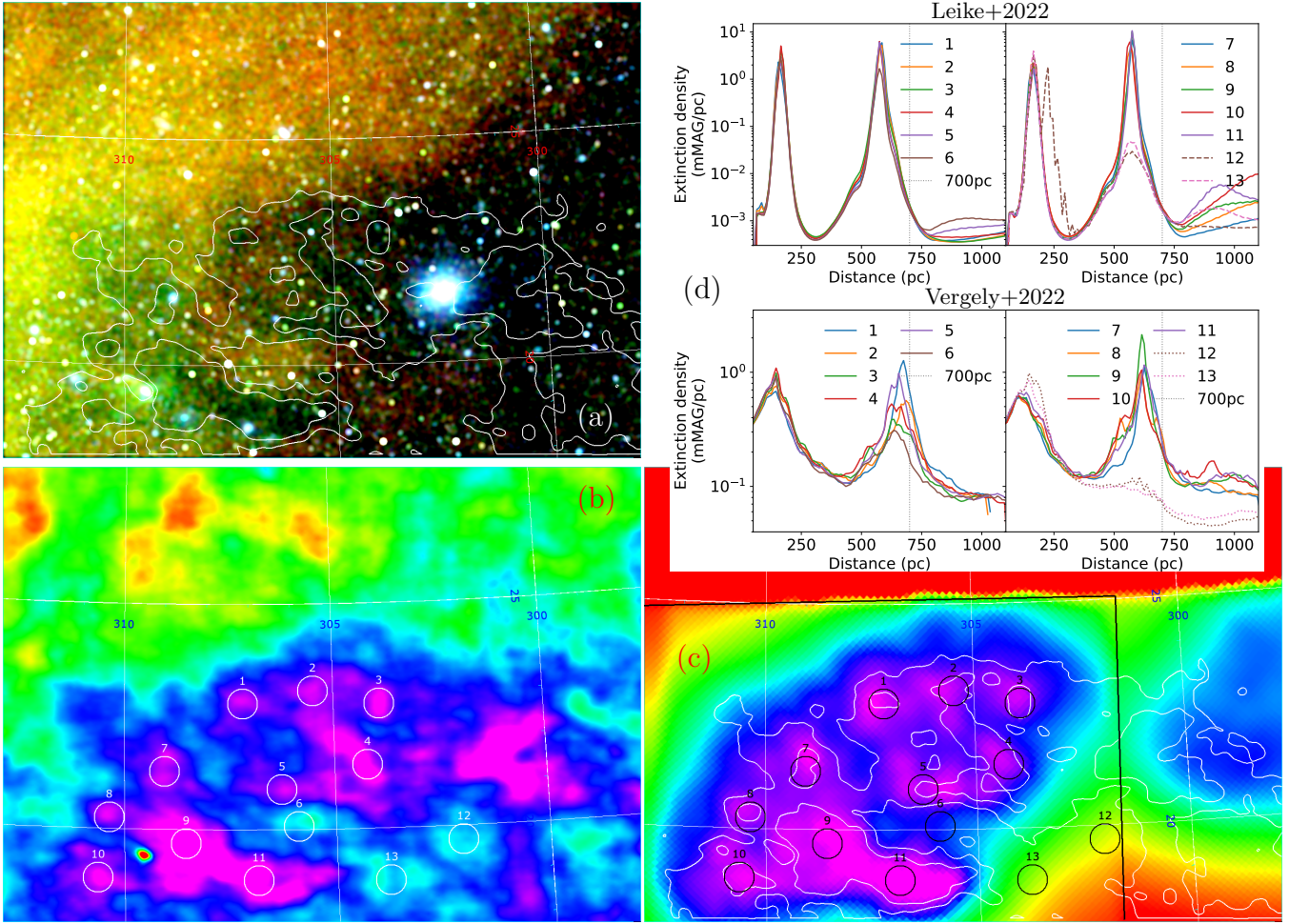

Fig.~\ref{fig:LP_images} compares three images in the G308+21 region:
the \citetalias{Le22} extinction map integrated in the distance range 520--650~pc,
the total \NH map,
and the X-ray image extracted from the five eROSITA all-sky surveys \citep[eRASS:5;][]{Predehl2021,Merloni2024} using the eROSITA Science Analysis Software System \citep[eSASS;][]{Brunner2022}.
The G308+21 cloud, which appears as over-densities in the Av map and \NH map, also appears as an X-ray shadow below (south of) the LPC.
The total \NH map matches the X-ray shadows (darker or bluer regions) perfectly, suggesting that the X-ray-emitting cloud is located beyond all the obscuring material.
The G308+21 cloud falls in a single patch of the \citetalias{Le22} data and thus a high-quality \citetalias{Le22} extinction map was obtained in this region, integrating the extinction in the distance range 520--650~pc (panel (c) of Fig.~\ref{fig:LP_images}).
The extinction map extracted from the \citetalias{Ve22} dust cube is very similar, although with a slightly lower resolution.
Despite contamination of the Galactic plane layer, the shape of G308+21 on the extinction map matches the high-resolution total \NH map very well.
Therefore, G308+21 must obscure the X-ray emissions of the LPC.

We selected $13$ positions within this region and extracted the dust distance profile at each position.
The \citetalias{Ve22} and \citetalias{Le22} dust cubes show similar distance profiles.
As displayed in Fig.~\ref{fig:LP_images}, there is always a peak below 200pc, which is due to the thin layer of material ( $<$ 200pc) in the Galactic plane.
A second peak of around 600pc is prominent inside G308+21 (regions 1--11, solid lines).
For comparison, this peak recedes significantly in regions 12 and 13 ( dashed lines), which are outside the core region of G308+21.
Because of the different data sets and methods used by \citetalias{Ve22} and \citetalias{Ve22}, the distance measured by the latter is larger.
The peak distance of G308+21 is $\sim$650pc due to \citetalias{Ve22} and $\sim$570pc due to \citetalias{Le22}.
With a projected angular scale of $\sim 10\arcdeg$, at a distance of 650pc, the size of G308+21 would be $\sim$110pc.
Based on these results, we present a robust lower limit of 700pc to the distance of the LPC.

\subsection{The distance to the root of the NPS}
\begin{figure*}[hptb]
\centering
\parbox{0.7\textwidth}{
\includegraphics[width=0.35\textwidth]{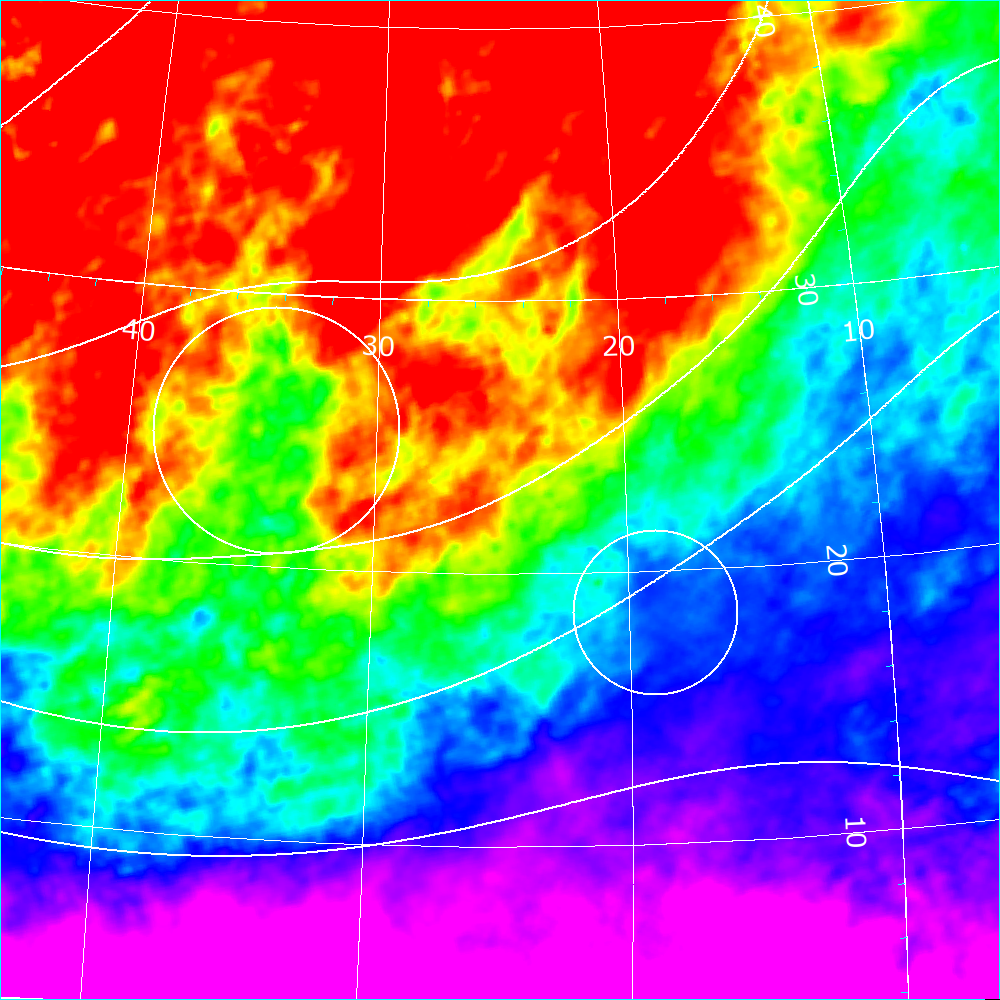}
\includegraphics[width=0.35\textwidth]{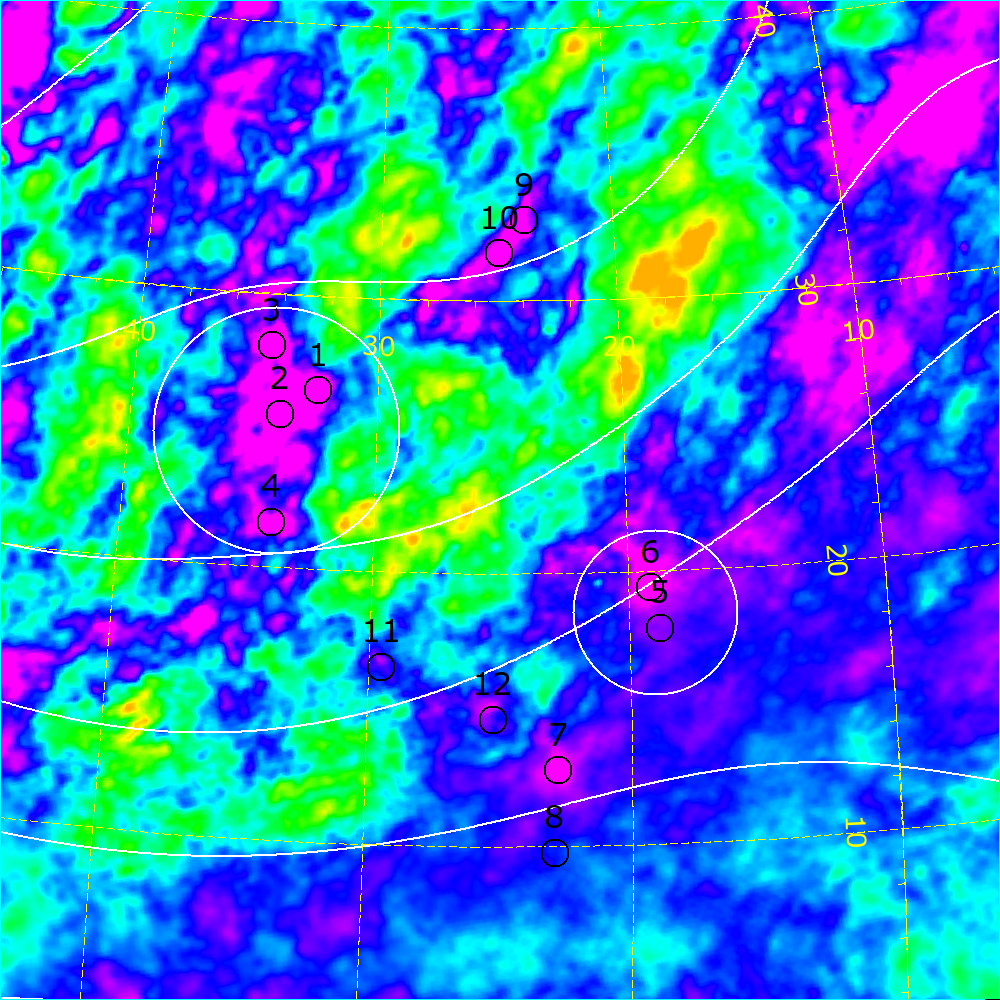}\\
\includegraphics[width=0.35\textwidth]{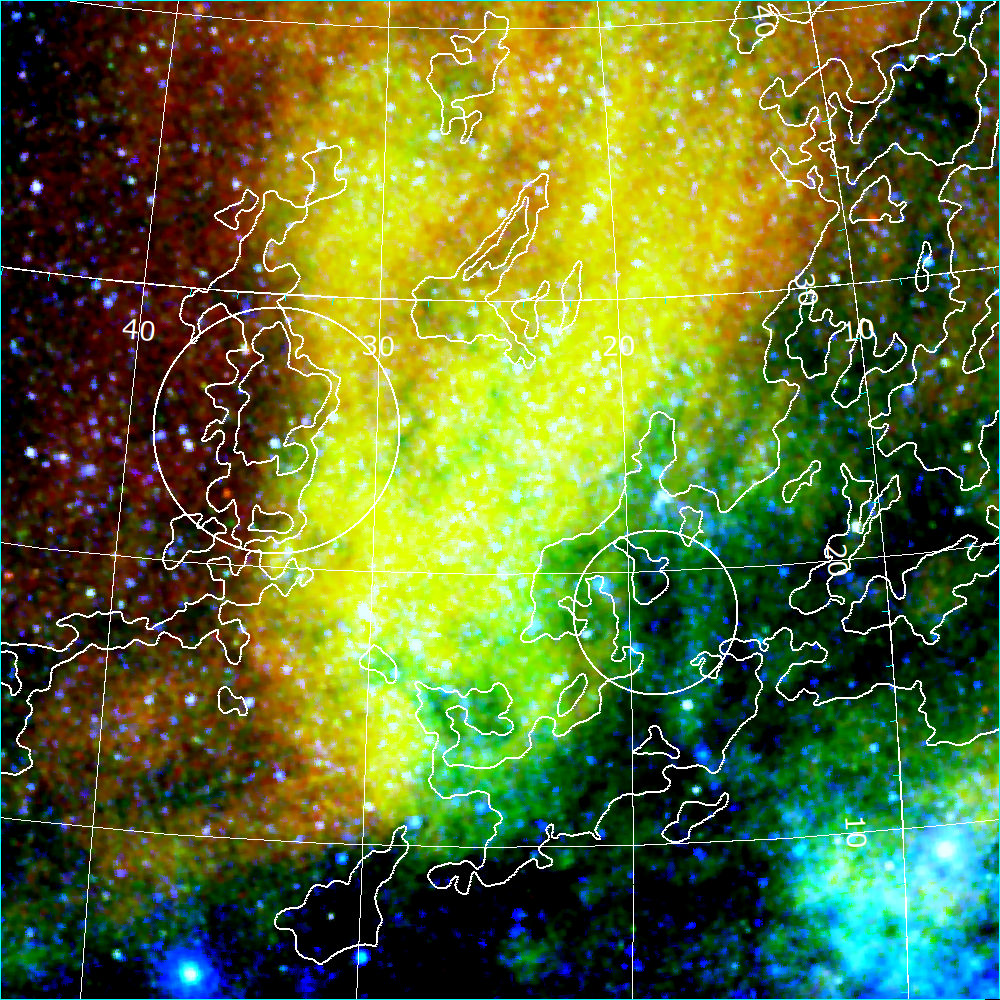}
\includegraphics[width=0.35\textwidth]{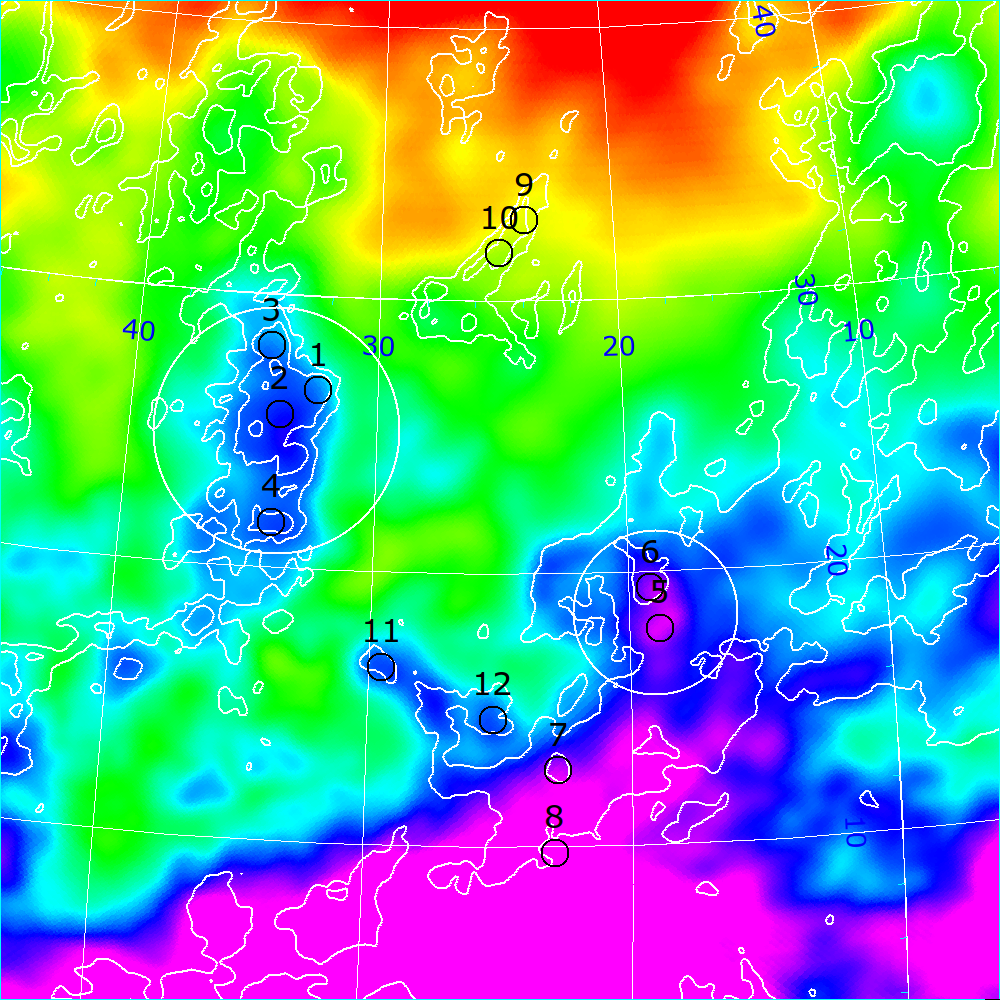}
}
\hspace{0.1cm}
\parbox{0.28\textwidth}{
\begin{overpic}[scale=0.31]{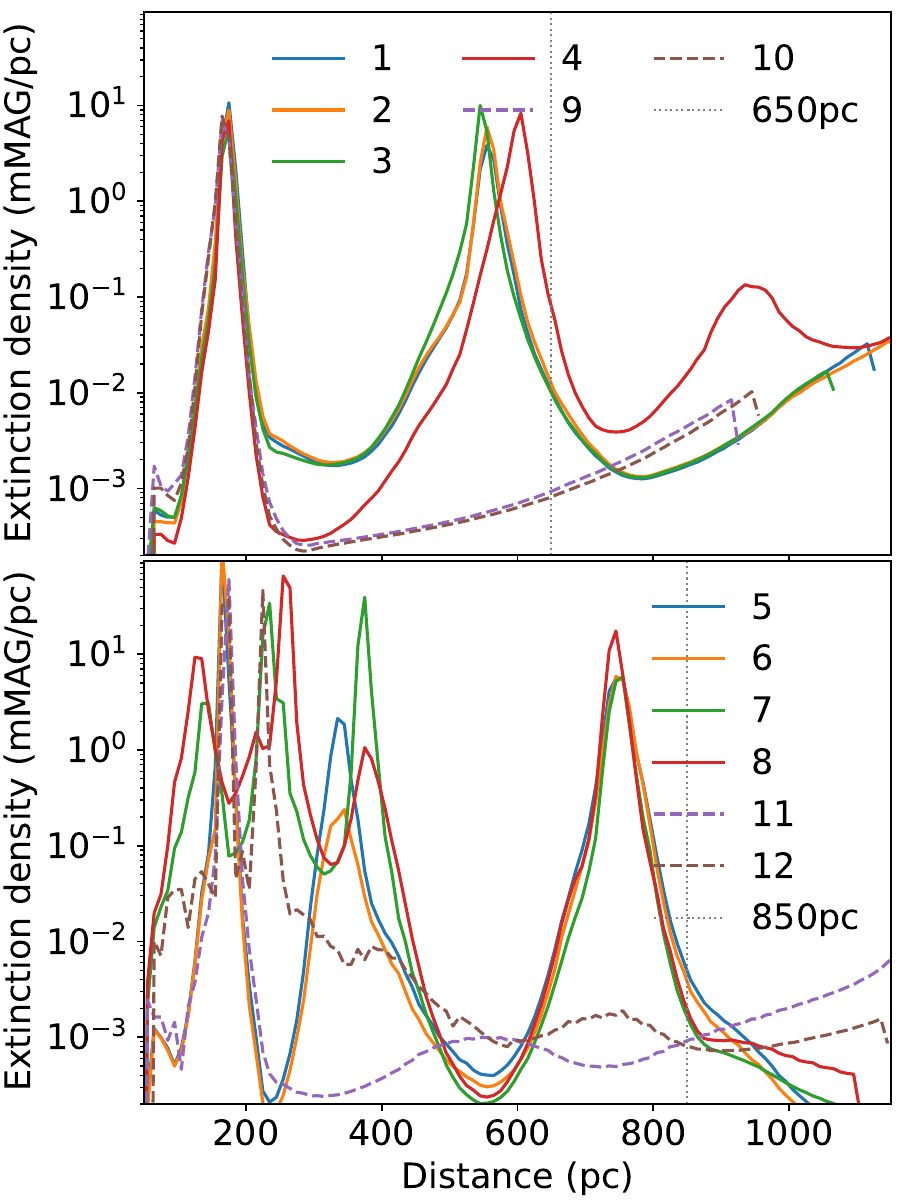}
\end{overpic}
\begin{overpic}[scale=0.31]{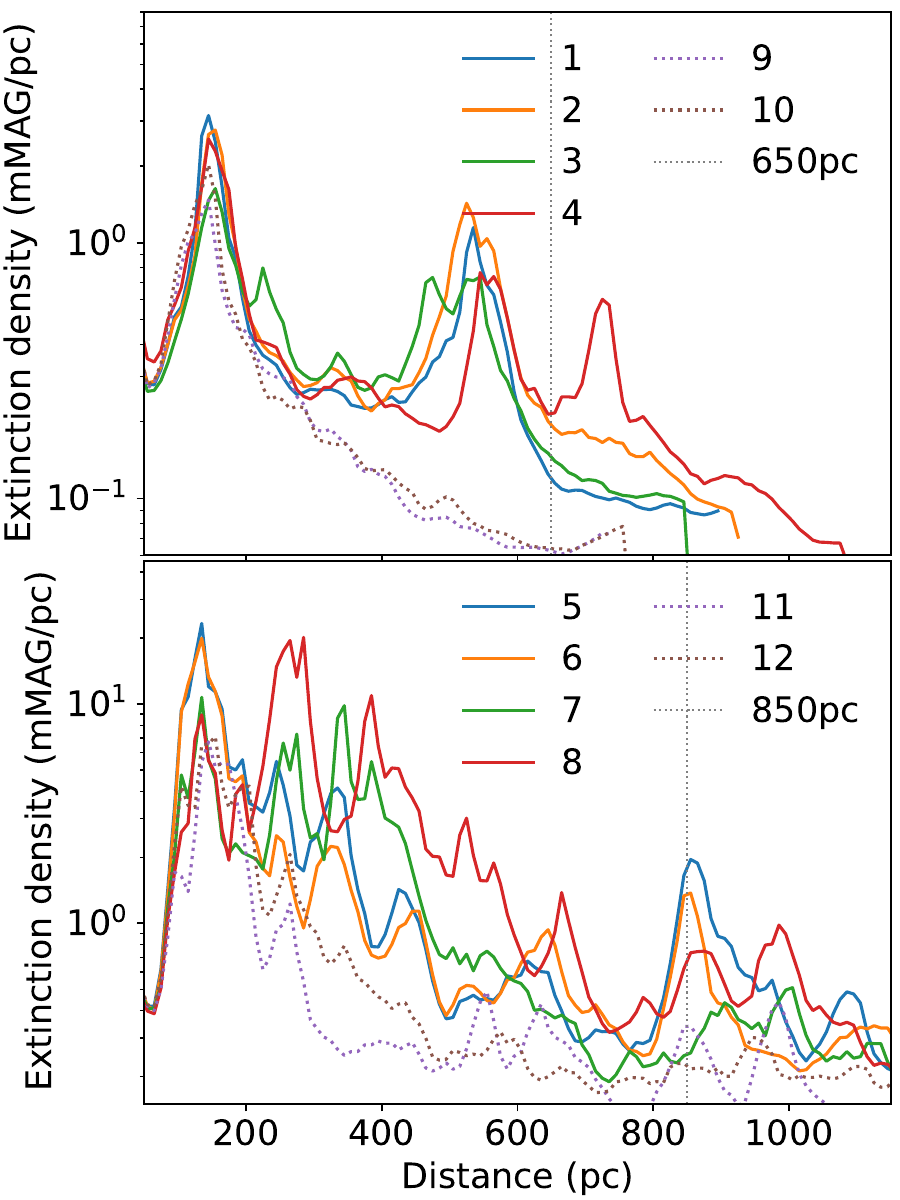}
  \put(-117,105){\color{white}\Large(a)}
  \put(-102,105){\color{white}\Large(b)}
  \put(-117,93){\color{white}\Large(c)}
  \put(-102,93){\color{blue}\Large(d)}
  \put(3,100){\color{black}\Large(e)}
\end{overpic}
}
\caption{
  Panel (c) displays the eROSITA X-ray RGB image published in \citet{Predehl2020}.
  Panel (a) and (b) display the total\NH map and the high frequency \NH map, which is calculated by subtracting the low frequency map (white contours) from the total-\NH map.
  %The contours of the \citetalias{Le22} dust extinction map in the distance range 540--640pc (white) and 650--850pc (yellow) are overplotted in the high-frequency \NH map.
  Panel (d) displays the \citetalias{Ve22} dust extinction map in the distance range 400--1000pc.
  %All four images are plotted in the Galactic system with a center of 25.139\arcdeg, $22.761\arcdeg$, and a size of 36.3$\times$36.3\arcdeg.
  The regions of G33+25 and G19+18 are marked with two big white circles.
  All the extinction and \NH maps use a rainbow color map, where purple indicates large values.
  The contours of the high-frequency \NH map are overplotted on the X-ray image and the extinction map in white, and in the former case, we removed some small patches of the contours for the clarity of the X-ray image.
  Panel (e) displays the \citetalias{Le22} (upper) and \citetalias{Ve22} (lower) distance profiles of extinction in the $12$ positions that are marked on the \NH and extinction maps with 1\arcdeg-diameter circles, separating the positions into two groups for clarity.
  \label{fig:NPS_images}}
\end{figure*}

%The shape of NPS's southern terminus matches the high-\NH region well, confirming that the southern terminus is absorption bounded \citep{Lallement2016}.

In Fig.~\ref{fig:NPS_images},
panel (c) displays the eROSITA X-ray RGB image \citep[from][]{Predehl2020} of the root of the NPS, and panel (a) displays the total \NH in the same region.
Due to the low latitude near the gas-rich Galactic plane, the total \NH map shows a large gradient.
Using a low-pass filter in the Fourier domain, we extracted a low-frequency component of the total \NH map (white contour overplotted on the \NH map), and then subtracted it to make a high-frequency \NH map (panel (b)), in which the shapes of the G19+18 cloud and \edit1{part of the G33+25} clouds become clearer.
We plot the contours of the high-frequency \NH map on the X-ray image and the \citetalias{Ve22} dust extinction map within 400--1000pc (panel (d)).
The G33+25 cloud shows very similar shapes on the \NH and Av maps, and its western border matches the X-ray shadow on the NPS very well.
The G19+18 cloud shows an elongated shape in the north-south direction on the Av map, which perfectly matches the elongated X-ray shadow.
We could only plot the contour of its northern part on the high frequency \NH map, because its southern part is overwhelmed by the thick foreground layer of material.
We conclude that these two dusty clouds are obscuring the NPS.

% Slightly different from regions 1--3, the dust in region 4 extends to a larger distance (another peak at 725pc).
% Region 9, appearing as a tail of G33+25 in the 2D map, does not colocate with G33+25 in terms of distance.
%Regions 5--8 compose G19+18, which is represented by the peak at 850pc.
%The obscuration at the positions of regions 10--13 is due to less-distant structures inside the Galactic plane and thus not of interest.

As displayed in Fig.~\ref{fig:NPS_images}, we also extracted dust distance profiles at a few positions in these regions from the \citetalias{Le22} and \citetalias{Ve22} dust cubes. Regions 1--4, which are chosen within G33+25, show a second peak between around 550~pc.
For comparisons, regions 9, 10, 11, and 12 mark two elongated absorption (\NH overdensity) features that also perfectly match X-ray shadows. According to their distance profiles, they are both local features near the Galactic plane.
The G19+18 cloud is marked by regions 5 and 6, whose distance profiles show identical distant peaks, although with a difference in distance measurements between the two versions of dust cubes. In \citetalias{Ve22}, this G19+18 peak centers at $\sim$850pc and extends to 1kpc. In \citetalias{Le22}, this peak centers at $\sim$750pc and extends to $\sim$850pc.
Regions 7 and 8 are chosen within the Aquila-Serpens molecular cloud, which obscures the southern terminus of the NPS at $b<10\arcdeg$ \citep[the triangle purple region at the bottom of panel (d);][]{Sofue2015}. In \citetalias{Le22}, regions 7 and 8 share the same distance peak as G19+18, indicating that G19+18 is a high-latitude extension of the Aquila-Serpens cloud.
The distance peaks of regions 7 and 8 are not detected by \citetalias{Ve22} as clearly as by \citetalias{Le22}.
Based on these results, we conclude that at an intermediate latitude ($\sim 20\arcdeg$), the NPS has a robust distance lower limit of 850pc.

\section{The Shape of the eROSITA Bubbles}
\label{sec:shape}

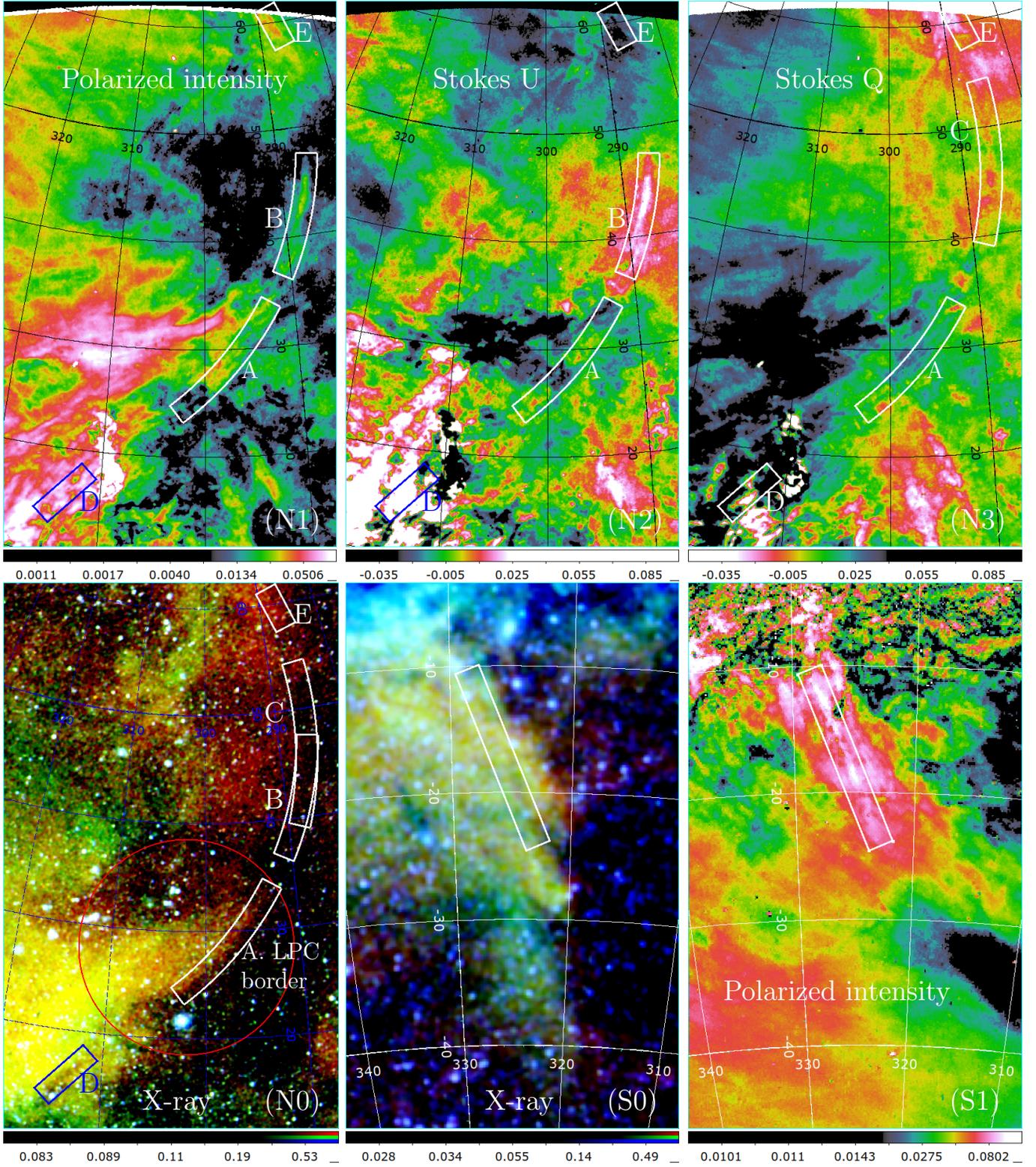
\begin{figure*}[hptb]
  \centering
  \begin{overpic}[scale=0.31,valign=c]{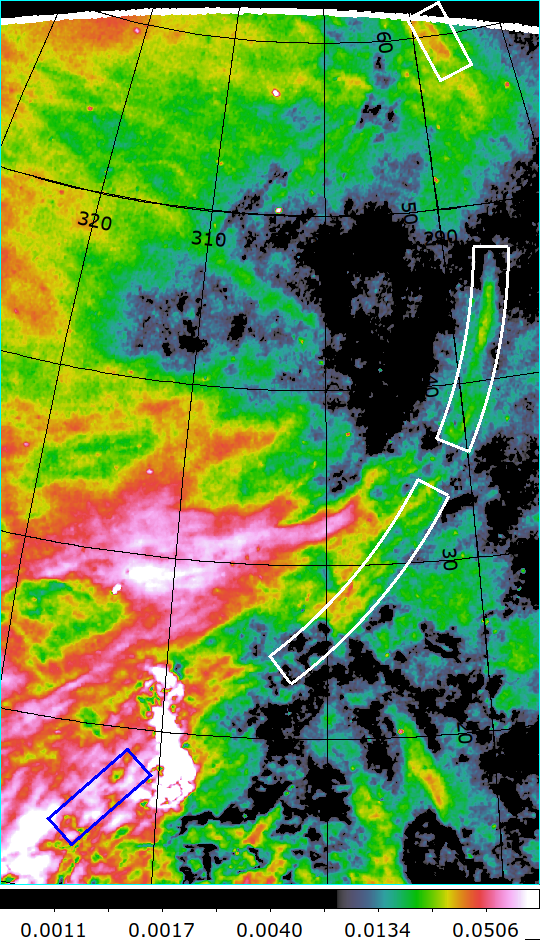}
    \put(45,9){\color{white}\Large(N1)}
    \put(10,85){\color{white}\Large Polarized intensity}
    \put(41,35){\color{white}\large A}
    \put(45,61){\color{white}\Large B}
    \put(13,12){\color{blue}\Large D}
    \put(50,93){\color{white}\Large E}
  \end{overpic}
  \begin{overpic}[scale=0.31]{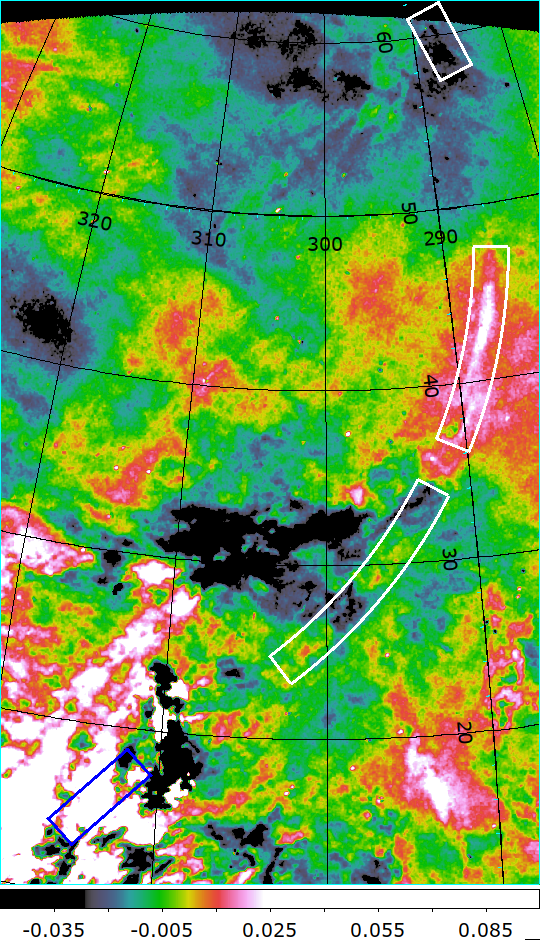}
    \put(45,9){\color{white}\Large(N2)}
    \put(15,85){\color{white}\Large Stokes U}
    \put(41,35){\color{white}\large A}
    \put(45,61){\color{white}\Large B}
    \put(13,12){\color{blue}\Large D}
    \put(50,93){\color{white}\Large E}
  \end{overpic}
  \begin{overpic}[scale=0.31]{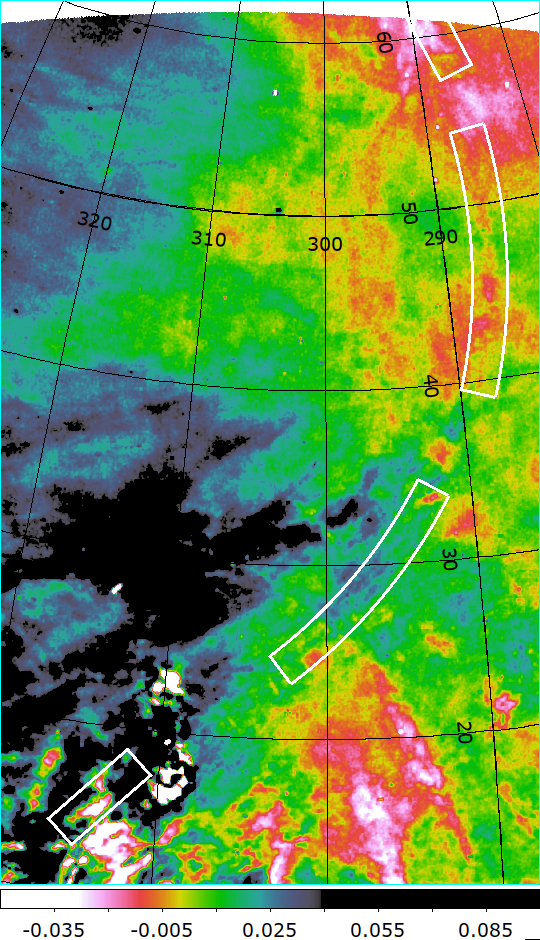}
    \put(45,9){\color{white}\Large(N3)}
    \put(15,85){\color{white}\Large Stokes Q}
    \put(41,35){\color{white}\large A}
    \put(45,76){\color{white}\Large C}
    \put(13,12){\color{white}\Large D}
    \put(50,93){\color{white}\Large E}
  \end{overpic}
  \begin{overpic}[scale=0.31]{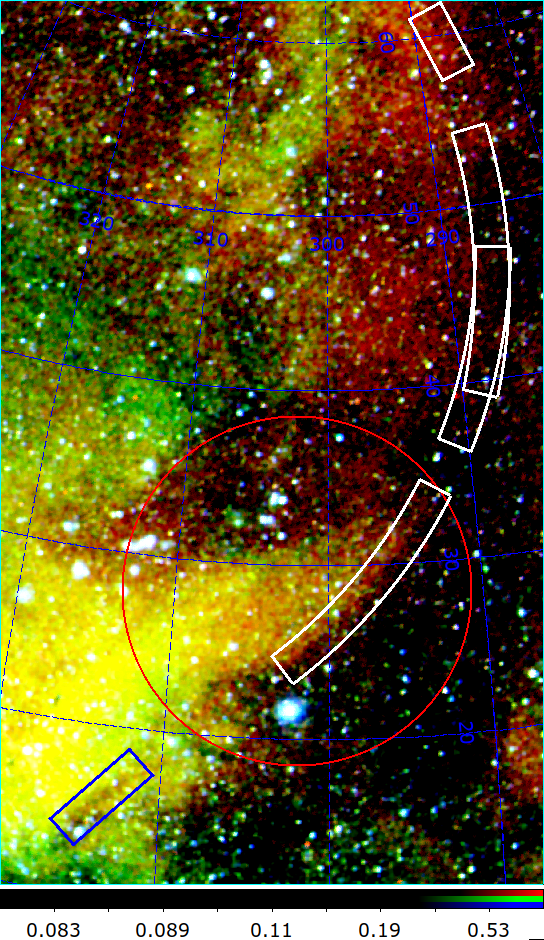}
    \put(45,9){\color{white}\Large(N0)}
    \put(41,35){\color{white}\large A. LPC}
    \put(45,61){\color{white}\Large B}
    \put(45,76){\color{white}\Large C}
    \put(13,12){\color{blue}\Large D}
    \put(50,93){\color{white}\Large E}
    \put(41,30){\color{white}\large border}
    \put(24,9){\color{white}\Large X-ray}
  \end{overpic}
  \begin{overpic}[scale=0.31]{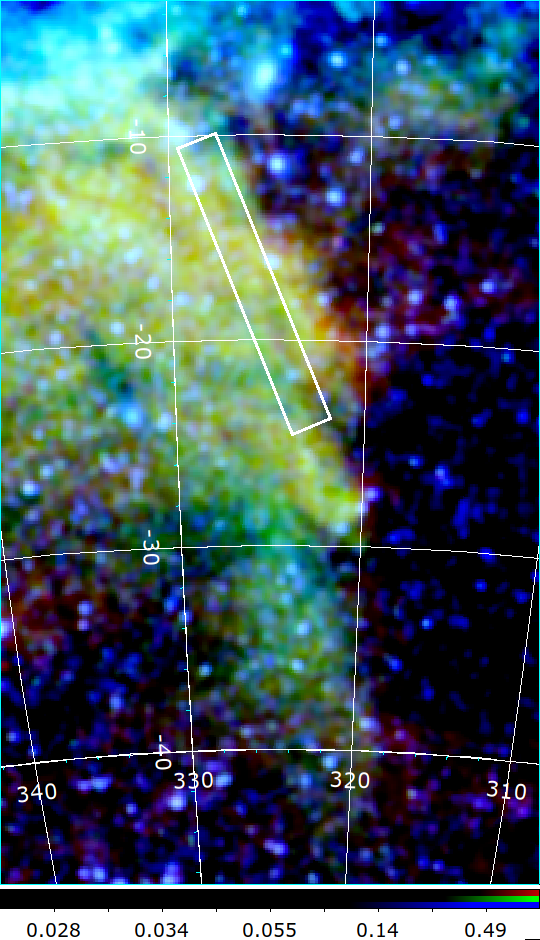}
    \put(45,9){\color{white}\Large(S0)}
    \put(24,9){\color{white}\Large X-ray}
  \end{overpic}
  \begin{overpic}[scale=0.31]{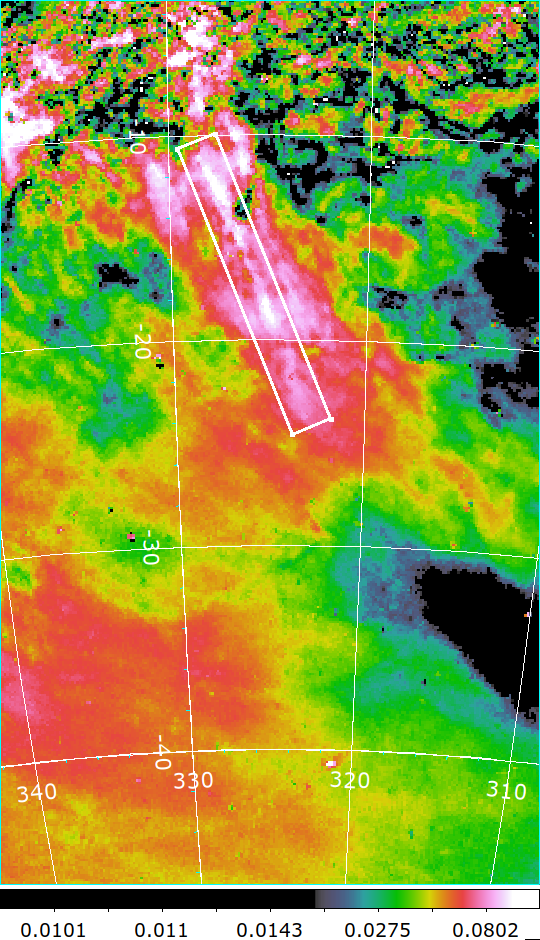}
    \put(45,9){\color{white}\Large(S1)}
    \put(6,28){\color{white}\Large Polarized intensity}
  \end{overpic}
  \caption{
    Panels (N0), (N1), (N2) and (N3) display the northwestern bubble border in terms of the X-ray RGB image \citep{Predehl2020}, the SPASS polarized intensity map, and Stokes U and Q maps.
    Five sections of the border in this field are marked in white or blue strips (A, B, C, D, and E) with a width of 2\arcdeg.
    The LPC is marked by a red circle in panel (N0).
    %plotted in the Galactic system with a center of $304\arcdeg$, $37\arcdeg$, and a size of $30.9\times 50.5\arcdeg$.
    Panels (S0) and (S1) display the southwestern bright rim in terms of X-ray RGB image \citep{Predehl2020} and the SPASS polarized intensity map.
    The position of the radio rim is marked in a white strip with a width of 2\arcdeg.
    %plotted in the Galactic system, centered at $325\arcdeg$, $-25\arcdeg$ and with a size of $26.3\times 42.9\arcdeg$.
    \label{fig:X_I_U_Q}}
\end{figure*}

\subsection{The Border of the eROSITA Bubbles}
\label{sec:border}
%Over 70\% SNRs in the Milky Way have a shell-type morphology characterized by a limb-brightened rim or shell at the location of the shock front, which usually coincides with the onset of X-ray emissions coming from the interior of the bubble. Radio rims represent synchrotron emissions on the shock front, where the particles are accelerated and the magnetic field is compressed with possibly other mechanisms in action to amplify it \citep[see reviews by][]{Reynolds2012,Dubner2017}. The synchrotron emission is expected to be linearly polarized.
%The polarized fraction is generally found to be $10\sim15\%$ for young SNRs. (maybe EB is old)
Assuming that the eROSITA bubbles are blown by energy injection from the GC, although with a different central engine compared to SNR, the physics at the shock front could be similar, i.e., showing a synchrotron rim at the periphery and X-ray emissions interior of it.
The S-band Polarization All-Sky Survey \citep[SPASS,][]{Carretti2019}, although covering only the south equatorial hemisphere, provides the best opportunity to reveal polarized radio rims because it was performed at 2.3~GHz, a frequency that is high enough to avoid strong depolarization and low enough to retain a high S/N.

Fig.~\ref{fig:X_I_U_Q} displays the X-ray and SPASS polarized maps in a region in the northwest sky around the LPC.
We found a few segments of arcs that trace the border of the eROSITA bubble.
In region A, the X-ray emitting LPC shows a round border in the latitude range $24\sim 33\arcdeg$.
In region B, where the X-rays fade, a prominent arc appears in the Stokes U map ($37\sim47\arcdeg$). At higher latitudes, this radio arc disappears in Stokes U but extends in Stokes Q ($39\sim 54\arcdeg$, region C).
The X-ray border and the radio arcs are perfectly in line. We attribute them to the shock front \edit1{and thus the outer border} of the northern bubble.
\edit1{
  We remark that, with this definition of the bubble's outer border, the bright tail of the NPS, i.e., the green feaure in Fig.~\ref{fig:X_I_U_Q} panel (N0) extending from the top to $b\sim 40\arcdeg$ at a longitude of 300$\sim$310\arcdeg, is located significantly inward of the border.
  }

At a higher Galactic latitude ($l=287.4\arcdeg$, $b=60.2\arcdeg$) in region E near the celestial equator, a short arc appears in both Stokes U and Q, although contaminated by nearby features.
At a lower Galactic latitude ($l=314\arcdeg$, $b=15\arcdeg$) in region D, the X-ray image shows a dark slit, which cannot be explained by any obscuration feature at this position in the total \NH map.
Across the slit, the X-ray brightness changes significantly. Also, there is likely a corresponding radio rim on the Stokes Q map. We also consider them as the shock front of the northern eROSITA bubble.

The southern eROSITA bubble \edit1{is faint in X-ray emission and only shows a bright rim} at low latitudes in the southwestern quadrant. As shown in Fig.~\ref{fig:X_I_U_Q} panel (S0), the X-ray border is not clear in this region.
In the polarized radio intensity map (panel (S1)), we noticed a long rim at this position \edit1{(marked with a white stripe), although with some indistinct short features around it.}
We also consider this radio feature to be possibly the shock front.

We combined these shock front \edit1{arcs} with the apparent X-ray outer border on the eROSITA all-sky map \citep{Predehl2020} to define the global border of the eROSITA bubbles and attempted to explain the border shape with a 3D model in the next section.

\subsection{Modeling the bubbles as containers of hot gas}
\label{sec:model}
\begin{figure*}[hptb]
\centering
\begin{minipage}{0.20\textwidth}
\vspace{1cm}
\includegraphics[width=\textwidth,right,valign=c]{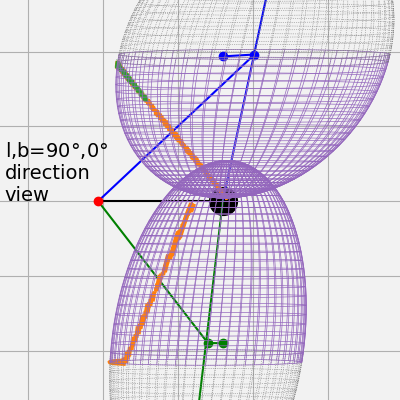}
\vspace{1cm}
\includegraphics[width=\textwidth,right,valign=c]{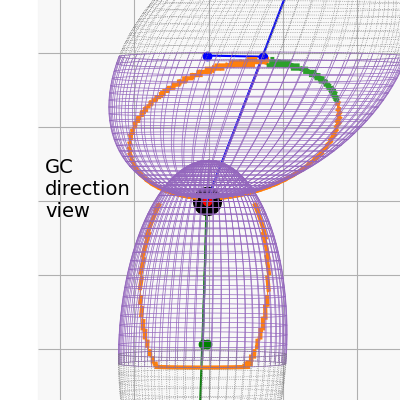}
\end{minipage}
\begin{minipage}{0.78\textwidth}
  \begin{overpic}[scale=0.45,valign=c]{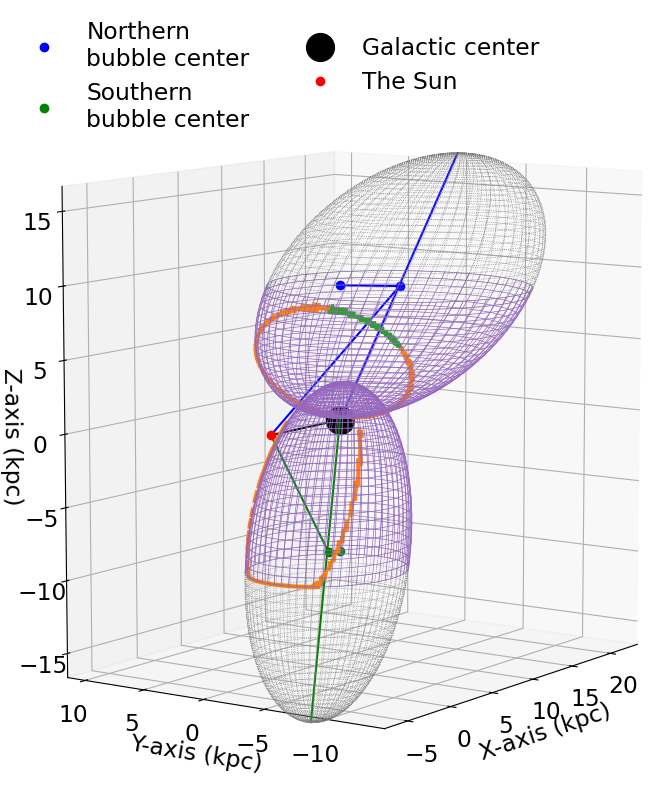}
  \end{overpic}
  \begin{overpic}[scale=0.40,left,valign=c]{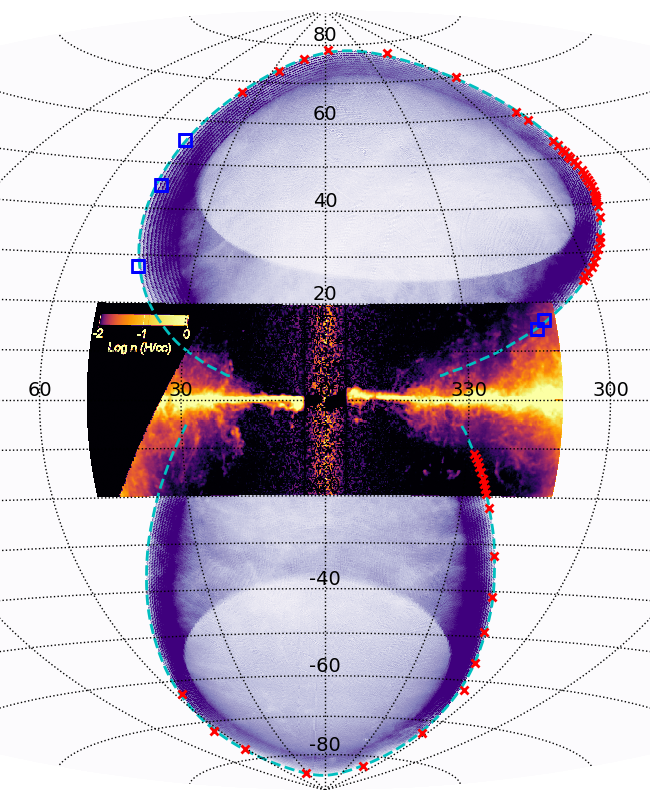}
    \put(-80,29){\color{violet}\Large(a)}
    \put(-80,24){\color{violet}\Large(b)}
    \put(-9,53){\color{violet}\Large(c)}
    \put(1,53){\color{violet}\Large(d)}
  \end{overpic}
\end{minipage}
\caption{ Panel (c) displays the 3D skewed cup model (purple), created by cutting off the high-$Z$ part (grey) of the skewed ellipsoid model.  The blue and green points indicate the centers of the ellipsoids before and after shifting.  The red and black points indicate the locations of the Sun and the GC.  The LOS tangent of the 3D surface is marked in orange and green, and green indicates the $l>180\arcdeg, 35\arcdeg<b<65\arcdeg$ section, where the border is determined by the radio arcs shown in Fig.~\ref{fig:X_I_U_Q}.  Panels (a) and (b) display the view of the 3D model from two aspects facing east ($l=90\arcdeg$) and the GC, respectively.  Panel (d) displays the Hammer-aitoff projection of the skewed cup model.  The red cross points and blue square points are manually selected marks of the border position (see also Fig.~\ref{fig:Fig1}).  Only the red points were used to constrain the 3D model, whose border is displayed by the cyan dashed lines.  The H{\scriptsize I} column density map along the Galactic tangent circle presented by \citet{Sofue2017} is overplotted on the 2D map. }
    \label{fig:SkewCup}
\end{figure*}

We explain the asymmetric shape of the bubble border defined above with the simplest geometric model, i.e., an ellipsoid:
\begin{equation}
  \frac{(X-C_X)^2}{R_X^2}+\frac{(Y-C_Y)^2}{R_Y^2}+\frac{(Z-C_Z)^2}{R_Z^2}=1,
\end{equation}
where $(C_X,C_Y,C_Z)$ is the center position and $R_X,R_Y,R_Z$ are the radii in the three axes.
We define the X-axis as the direction to GC, the Y-axis as the direction to the east ($l=90\arcdeg$), and the Z-axis to the north.
Initially, we set $C_X$ to the solar Galactocentric distance \citep[8~kpc,][]{Reid1993} and $C_Y$ to 0, so that the ellipsoid heads straight upward or downward.
To explain the east-west asymmetry, we skewed the ellipsoid by shifting each point in the XY plane as a linear function of Z:
\begin{equation}
  X=X+\alpha_XZ,\ \ Y=Y+\alpha_YZ,
\end{equation}
where $\alpha_X$ and $\alpha_Y$ are the skew factors.
Since we only focused on the border of the eROSITA bubbles, we assumed a thin shell that only emits uniformly on the surface, which gives rise to a projected 2D shell because of limb brightening.
\edit1{The 2D border corresponds to the line-of-sight (LOS) tangent of the 3D surface.}
By minimizing\footnote{We used the \texttt{scipy.optimize} package.} the distance between the projected \edit1{model shell to the border positions determined by X-ray and radio data}, the model parameters can be constrained.
Considering that the gas density could be much lower farther away from the Galactic plane, the bubble does not have to be closed at large $Z$.
As illustrated in the review of \citet{Lallement22}, the X-ray-emitting region might look like a cup or bowl on the Galactic plane. 
Adding a cutoff height parameter $Z_H$, we turned the ellipsoid model into a cup model by removing the part above $Z_H$ and then fitted this cup model to the border positions.
The best-fit 3D cup model is shown in Fig.~\ref{fig:SkewCup}.
For presentation purposes, we projected $11$ 3D thin shells into 2D space with identical surface brightness, which have radii and $Z_H$ between 100\% and 90\% of the best-fit values with a 1\% step.
The outermost shell corresponds to our best-fit model and determines the projected 2D border.
We also made absorption corrections to this 2D map using the HI4PI \NH map, assuming a 0.2~keV hot plasma model.

In this work, we focus mainly on the northern bubble, whose border is accurately defined in the five regions A--E in Fig.~\ref{fig:X_I_U_Q}.
\edit1{Along the LPC border in region A and the radio arcs in region B, C, and E, we manually drew $33$ points to trace the border, as shown by the red cross points in panel (d) of Fig.~\ref{fig:SkewCup}.}
For the NPS, the X-ray outer border is clear at $b>65\arcdeg$ (northern border) but unclear at $b<65\arcdeg$ (eastern border), because of some faint emissions that envelope NPS's brightest part at larger longitude with a width of $\lesssim 10\arcdeg$.
\edit1{As shown in Fig.~\ref{fig:Fig1} and Fig.~\ref{fig:SkewCup}, we drew six red points to trace the clear northern border and three blue points to mark the indistinct eastern border.}
The short rim in region D is not clear either \edit1{as the outer border of X-ray emissions; thus, we drew two blue points to mark this rim.
We excluded such unclear regions (blue points) and used only the reliable border positions (red cross points) to constrain the 3D model.
As shown in Fig.~\ref{fig:SkewCup},  we found that the border of the best-fit model not only describes the red points perfectly but also matches the blue points well}, suggesting that the blue points are also reasonable choices as the border of the northern eROSITA bubble.

\edit1{It has been noticed that the H$_\footnotesize\textrm{I}$ voids around the GC compose a bipolar conical cavity approximately coincident in position with the eROSITA bubbles \citep{Lockman2016,Sofue2017,Sofue2021}.
 \citet{Sofue2017} calculated the H$_\footnotesize\textrm{I}$ density along the Galactic tangent circle, which is a curve in the Galactic plane crossing the GC and with a distance to the Sun of $R_0 \cos l$, where $R_0$ is the solar Galactocentric distance and $l$ is the Galactic longitude.
 We overplotted this H$_\footnotesize\textrm{I}$ distribution in Fig.~\ref{fig:SkewCup} panel (d).
  In the northwestern sky at $b<20\arcdeg$, our best-fit model crosses the rim D (two blue points) and extends along the wall of the H$_\footnotesize\textrm{I}$ conical cavity. Again, this suggests that the rim D is a reasonable choice as the bubble border.
There is X-ray emission to the west of rim D and thus outside our best-fit model (see Fig.\ref{fig:X_I_U_Q} panel (N0)):
we cannot}
rule out the possibility that this X-ray emission is also due to the northern bubble. Irregular convex \edit1{structures} on the bubble surface at low latitudes are possible because the gas density near the Galactic plane could be highly non-uniform.

\edit1{In summary, we find that the best-fit northern bubble is approximately a 9~kpc radius spherical bubble (before shifting), with $R_X$, $R_Y$, $R_Z$, and $C_Z$ of 8.6, 9.3, 9.1, and 9.3~kpc. This is larger than the \citet{Predehl2020} spherical model (which has a radius of 7~kpc) because, before matching to the 2D projection, it is skewed by shifting the $X,Y$ coordinates away from the Sun with the skew factors $\alpha_X=0.22, \alpha_Y=-0.38$.
  We remark that the 3D size measured through the 2D projection is proportional to the solar Galactocentric distance, which in this work was assumed to be 8~kpc.
  For the cutoff height $Z_H$, we obtained a value of 9~kpc, which is approximately the highest point of the LOS tangent of the 3D ellipsoid surface (marked in orange and green in Fig.~\ref{fig:SkewCup}). In this case, the obtained $Z_H$ value is only a lower limit, as any higher $Z_H$ has no impact on the LOS tangent.}

For the southern bubble, 
\edit1{the border is not as distinct as in the north, because of fainter X-ray emission and complex radio features}, but the manually chosen border positions are also well described by the same 3D cup model.
\edit1{The best-fit model has $R_X$, $R_Y$, $R_Z$, and $C_Z$ of 6.2, 5.4, 12, and -10~kpc, and has skew factors $\alpha_X=0.11, \alpha_Y=-0.03$, being more elongated and less inclined than the northern bubble.
The high-$Z$ cutoff parameter $Z_H$ is -9.4~kpc, which} took effect in the model by slightly changing the LOS tangent.
We consider this high-$Z$ cutoff as overfitting to the data. 
Since only the border position is taken into account, our model cannot provide robust evidence to distinguish the cases of an open cup \edit1{from} a closed bubble.
Using the closed-ellipsod model instead of the cup model, the best-fit parameters of the southern bubble remain similar.

\section{Discussion}
\label{sec:discussion}
\subsection{\edit1{The northern bubble composed of NPS and LPC}}
In \S~\ref{sec:clouds} we have argued that both the NPS's root ($>850$pc) and the LPC ($>700$pc) are distant at intermediate latitudes.
However, it is uncertain whether they compose two independent structures in 3D space or, instead, one single structure, i.e., the northern eROSITA bubble.
This appears to be an essential question about the definition of eROSITA bubbles.
The LPC reaches a maximum latitude of $33\arcdeg$, which corresponds to a height of 650 pc above the Galactic plane, if it is located at a distance of 1kpc.
In this case, the LPC is only an interesting structure at the scale of the Galactic plane (hundreds of pc).
In contrast, if they compose a single structure, this makes a compelling argument for the distant, giant bubble model.
Since the LPC and the NPS, separated by almost $90\arcdeg$, are both distant ($\sim 1$kpc) at intermediate latitudes, an individual structure made out of them should have a 3D center position that is more distant (plausiably a few kpc) and in the GC direction.
The most natural scheme of fitting such a structure into the Milky Way's morphology is a giant bubble rooted at the GC.

The radio arcs we found provide a connection between the NPS and the LPC. Considering the X-ray outer border and the radio arcs as proxies of the shock front, the high resolution of the eROSITA X-ray image and the SPASS radio image allows us to define the northern bubble's border accurately in three sections, the western border of the LPC \edit1{(region A in Fig.~\ref{fig:X_I_U_Q}), the northern border of the NPS (red points in Fig.~\ref{fig:Fig1}), and the radio arcs connecting them (regions B, C, and E in Fig~\ref{fig:X_I_U_Q}).}
Fitting the 3D cup model to these border positions, we found that the model not only describes these points perfectly, but also makes reasonable predictions of the bubble's border at positions where the X-ray outer border is not clear (blue points \edit1{tracing the eastern border of the NPS and rim D at around 310\arcdeg,15\arcdeg} in Fig.~\ref{fig:Fig1} and Fig.~\ref{fig:SkewCup}).
\edit1{Meanwhile, it also matches the shape of the H$_\footnotesize\textrm{I}$ conical cavity around the GC.}
Based on these findings, our definition of the northern bubble's border in terms of shock front is reasonable \edit1{and indicates} that the NPS and the LPC are both parts of the northern eROSITA bubble.

\subsection{The 3D cup model}
Although not closed, the 3D cup model can still be called a bubble model, as it describes the shock front of a powerful engine in the GC.
In the northern sky, since the obtained cutoff height $Z_H$ is not lower than the top of the tangent, the cup model and the closed ellipsoid model are equivalent, in the sense that they share the same projected (observable) border. 
However, we prefer the cup model because it predicts brighter X-ray emission at low latitudes and suggests more gas concentrated near the Galactic plane within a few kpc.
\edit1{This, in turn, opens up the discussion on the non-uniform gas distribution at the kpc scale. Dividing the 3D cup model into two parts by the LOS tangent, the observed X-ray emission is dominated by the side of the cup closer to the Sun.}
The root of the NPS is located around $l=27\pm10\arcdeg$, the direction to the near end of the Galactic bar or the head of the Scutum-Centaurus arm, and the LPC is located above the tangency of the Scutum-Centaurus arm \citep[$306\arcdeg$--$313\arcdeg$,][]{Churchwell2009}, suggesting that gas distribution related to the spiral structure of Milky Way could shape the appearance of the bubble in X-ray.
The 2D appearance of the eROSITA bubbles might then be the result of the expanding shock front (the cup) and its intercation with the 3D distribution of the hot gas inside the cup, particularly the gas heated \edit1{closer to the Sun}.
%A similar 3D model for the eROSITA bubbles has recently been proposed by Zhang et al. (submitted).

There may have been AGN activity in the GC a few million years ago, creating the Fermi bubbles \citep[see review by][]{Yang2018}.
\citet{Su2012} found a jet-like feature in the Fermi bubbles with a projected direction of $15\arcdeg$ from the north-south axis of the Galaxy.
Through the ionization cone of the GC AGN manifested on the Magellanic stream, \citet{BlandHawthorn2019} found a tilt direction broadly consistent with the Fermi bubble.
In this work, we explain the east-west asymmery of the northern bubble by tilting the bubble centered at the GC.
\edit1{If also attributing the eROSITA bubbles to a jet,}
our best-fit model predicts a northern jet direction of $l=299\arcdeg$, tilted from the north pole by $23.5\arcdeg$, which is also consistent with the Fermi-derived jet model of \citet{Su2012}.

However, after millions of years of travel, the shock front \edit1{might have forgotten the original jet direction and be shaped by the density gradient of the ambient medium} \citep{Sarkar2019,Sarkar2023}. The northern bubble tilting towards the west indicates that the gas density is lower on the west, allowing the shock front to move farther in this direction.
If this is true, the lower density might also explain why the X-ray emission fades in the northwest and only becomes bright again in the LPC, which is closer to the Galactic plane and thus likely has a higher density.

\edit1{
Compared with the northern bubble, the southern bubble not only is much fainter in X-rays, but also shows a different shape.
  As shown in Fig.~\ref{fig:SkewCup}, it is less inclined and tilted from the south pole only by a small angle of about $6.5\arcdeg$.
  It also has a more elongated shape, which extends to the north beyond the Galactic plane rather than roots at the GC,
  and it does not match the shape of the H$_\footnotesize\textrm{I}$ conical cavity at low latitudes.
  Possibly, because of the relatively lower density in the south sky, the southern bubble is less dissipated than in the north sky and therefore maintains the shape and direction of the jet more than the northern bubble \citep{Sarkar2019,Sarkar2023}.
  However, we cannot draw any conclusions on the basis of the bubble's border alone.
As shown in panel (a) of Fig.~\ref{fig:SkewCup}, the plane of the northern bubble's LOS tangent intersects the Galactic plane at the GC, but in the case of the southern bubble, the plane of the LOS tangent intersects the Galactic plane at a position closer to the Sun, resulting in a smaller fraction of the 3D surface in front of the LOS tangent.
Therefore, in this case, our ellipsoid model is less powerful in constraining the overall shape of the bubble, in the sense that the distant, undetected part of the southern bubble might not conform to this model. It is also possible that the ellipsoid model itself is valid for only a well-dissipated jet like the northern bubble but not for an elongated jet that dissipates only a small fraction of its energy into the ambient medium.
}

\section{Conclusions}
\label{sec:conclusion}
With the aim of investigating the nature of the eROSITA bubbles, we set apart a X-ray bright region in the northwest sky, the LPC, and studied its distance and connection with the NPS.

From the 3D dust distributions of the Milky Way, we found three isolated dusty clouds,
G33+25, G19+18 and G308+21, located near the NPS and the LPC.
The high-resolution total \NH maps and the dust extinction maps integrated in the corresponding distance ranges show highly consistent shapes of these clouds, and their shapes match the X-ray shadows very well.
From this we conclude that G33+25 and G19+18 obscure the root of the NPS, providing a distance lower limit of 850pc to it, and G308+21 obscures the LPC, providing a distance lower limit of 700pc to it.
In particular, these distance lower limits are measured at intermediate Galactic latitudes ($\sim 20\arcdeg$).
%rather than low latitudes ($<10\arcdeg$).

We found a few polarized radio arcs in the X-ray dark region between the NPS and the LPC, which smoothly connects the outer border of the two X-ray bright features.
Attributing the radio arcs to the limb-brightened shock front and the X-ray emissions to the hot gas interior of the shock front, we combined the radio arcs and the X-ray outer border to define accurately a new model for the border of the northern eROSITA bubble.
We found that the border defined in this way is well described by a 3D tilted ellipsoid or cup model.

\edit1{In summary, this work presents} the following findings:
\begin{enumerate}
    \item Like the NPS, the LPC's western border is sharp and round.
    \item Both NPS and LPC are obscured by \edit1{isolated clouds at known distances, and thus must be $\gtrsim$1~kpc from the Sun at intermediate latitudes.}
    \item The outer border of \edit1{the} NPS and LPC are smoothly connected by polarized radio arcs.
    \item The shape of the northern eROSITA bubble's border defined by the radio arcs and X-ray outer border can be attributed to \edit1{a} shock front and well described by a tilted ellipsoid model.
\end{enumerate}
Based on these findings, we conclude that, rather than being an independent feature, the LPC composes the northern eROSITA bubble together with the NPS. Considering that the two distant features (NPS and LPC) span an angle of $\sim90\arcdeg$, the bubble composed of them should be distant and giant with a scale of a few kpc. The most plausible model for such a structure is a giant bubble rooted in and blown by \edit1{episodes of energy release at} the Galactic Center.

\bibliography{lotus}

\begin{acknowledgements}
This work is supported by the National Natural Science Foundation of China NSFC-12393814. T.L. thanks Jun-Xian Wang, Zhen-Yi Cai, Guang-Xing Li, and Bing-Qiu Chen for insightful discussions.
This work is based on data from eROSITA, the soft X-ray instrument aboard SRG, a joint Russian-German science mission supported by the Russian Space Agency (Roskosmos), in the interests of the Russian Academy of Sciences represented by its Space Research Institute (IKI), and the Deutsches Zentrum für Luft- und Raumfahrt (DLR). The SRG spacecraft was built by Lavochkin Association (NPOL) and its subcontractors, and is operated by NPOL with support from the Max Planck Institute for Extraterrestrial Physics (MPE). The development and construction of the eROSITA X-ray instrument was led by MPE, with contributions from the Dr. Karl Remeis Observatory Bamberg \& ECAP (FAU Erlangen-Nuernberg), the University of Hamburg Observatory, the Leibniz Institute for Astrophysics Potsdam (AIP), and the Institute for Astronomy and Astrophysics of the University of Tübingen, with the support of DLR and the Max Planck Society. The Argelander Institute for Astronomy of the University of Bonn and the Ludwig Maximilians Universität Munich also participated in the science preparation for eROSITA.
The eROSITA data shown here were processed using the eSASS software system developed by the German eROSITA consortium.
G.P. acknowledges funding from the European Research Council (ERC) under the European Union’s Horizon 2020 research and innovation programme (grant agreement No 865637) and support from Bando per il Finanziamento della Ricerca Fondamentale 2022 dell'Istituto Nazionale di Astrofisica (INAF): GO Large program and from the Framework per l'Attrazione e il Rafforzamento delle Eccellenze (FARE) per la ricerca in Italia (R20L5S39T9).

\end{acknowledgements}

\end{document}